\documentclass[a4paper,11pt]{article}
\usepackage{jheppub} 
\usepackage{lineno}
\DeclareMathOperator{\Tr}{Tr}
\newcommand{\E}{\mathbb{E}}
\usepackage{subcaption}
\usepackage{float}
\usepackage{csquotes}
\newcommand{\llangle}{\langle\!\langle}
\newcommand{\rrangle}{\rangle\!\rangle}

\title{\boldmath Temporal correlations and chaos from spacetime kernels}

\author[a,b,c]{Rathindra Nath Das,}
\author[d]{Arnab Kundu,}
\author[e]{Matheus H. Martins Costa,}
\author[d]{Nemai Chandra Sarkar}

\affiliation[a]{Department of Particle Physics and Astrophysics, Weizmann Institute of Science, Rehovot 7610001, Israel}
\affiliation[b]{MIT Center for Theoretical Physics—a Leinweber Institute, Massachusetts Institute of Technology,
Cambridge, MA 02139, USA}
\affiliation[c]{Institute for Theoretical Physics and Astrophysics and Würzburg-Dresden Cluster of Excellence ct.qmat, Julius-Maximilians-Universität Würzburg, Am Hubland, 97074 W¨urzburg, Germany}
\affiliation[d]{Theory division, Saha Institute of Nuclear Physics, A CI of Homi Bhabha National Institute, 1/AF, Bidhannagar, Kolkata 700064, India}
\affiliation[e]{Institute for Theoretical Solid State Physics, IFW Dresden, Helmholtzstr. 20, 01069 Dresden, Germany}
\emailAdd{das.rathindranath@uni-wuerzburg.de, arnab.kundu@saha.ac.in, martinscostamh@gmail.com, nemaichandra.sarkar@saha.ac.in}

\abstract{We develop a finite-dimensional formulation of the recently introduced notion of “timelike entanglement”, defined in terms of two-point functions between operators supported on different Cauchy slices. Using a local orthonormal operator basis, we recast this construction in terms of a {\it generalized response tensor}. Building on this, we introduce a {\it generalized spacetime density kernel} (GSDK) corresponding to higher point correlation functions, including time-ordered as well as out-of-time-ordered correlators. We show that, the Haar-averaged $(2N)$-point function yields the $(2N)$-th moment of the spectral form factor (SFF), evaluated at an $N$-enhanced effective temperature. The correlation functions of the GSDK operators also yield the SFF, with an effective $(1/N)$-reduction of the physical time-scales. The GSDK places both scrambling diagnostics and spectral statistics on a similar footing and clarifies how higher-point correlators and non-trivial time ordering capture fine-grained dynamical information of a quantum system.}

\begin{document}
\maketitle
\flushbottom

\section{Introduction}

Quantifying quantum chaos has steadily evolved from the study of spectral statistics and semiclassical trace formulas to real-time probes of information flow, mixing, and operator growth. While entanglement entropy has proven to be an indispensable tool for characterising quantum systems, it is fundamentally a property of a state on a single spatial slice, offering a static snapshot of correlations \cite{Srednicki:1993im, Holzhey:1994we, Calabrese:2004eu}. A central challenge in understanding the chaotic dynamics in quantum systems is to develop a framework that moves beyond the static properties of quantum states to capture the structure of correlations in time. A complete understanding of phenomena such as thermalisation, information scrambling, and quantum chaos requires tools that are inherently dynamical.

Quantum chaos, in particular, manifests in distinct ways at different temporal scales, and a modern, information-theoretic picture of chaos must span both early-time Lyapunov growth and late-time random-matrix universality. At early times, its signature is the rapid delocalisation of quantum information, or scrambling, which is diagnosed by the exponential growth of out-of-time-order correlators (OTOCs) \cite{Larkin1969-wj}. The sharpest universal constraint on this growth is the Maldacena-Shenker-Stanford (MSS) bound on the Lyapunov exponent, $\lambda_L \le 2\pi T$, which is saturated in theories with gravity duals \cite{Maldacena:2015waa}. At late times, chaos is expected to exhibit universal features described by random matrix theory (RMT), where the fine-grained spectral statistics of the system's Hamiltonian govern the behaviour of observables like the spectral form factor (SFF), which displays a celebrated slope-dip-ramp-plateau structure \cite{mehta2004random, Bohigas:1983er, Haake:2010fgh}. A significant goal in the field is to bridge these early- and late-time descriptions. A crucial step in this direction was made in \cite{Cotler:2017jue}, where it was shown that suitable operator averages of OTOCs reproduce SFFs, thereby providing an explicit bridge between these viewpoints.

Despite this progress, most tools for chaos remain either spectral or correlational and typically state a property of a single time slice, such as entanglement entropy, or a specific correlator, such as OTOC. This motivates a framework that treats temporal correlation itself as the primary object and allows us to reason at the level of linear operators on tensor-product Hilbert spaces indexed by multiple times. Insights in this direction were provided by studies of the AdS/CFT, or holographic, duality pursued in \cite{Doi:2022iyj, Doi:2023zaf}. There it was shown that, much as the entanglement between complementary regions in a time slice of a quantum field theory (QFT) determines a corresponding spacelike area (and allows for the reconstruction of operators) in the bulk gravitational theory \cite{Ryu:2006bv, Penington:2019npb}, a certain analytical continuation of this real-space entanglement entropy is associated with a \textit{timelike} area in the bulk, with the dual subsystem in the boundary CFT having temporal extension and thus allowing for an investigation of correlations at different time slices.

From a purely holographic perspective, this ``timelike entanglement" suggests a way the time dimension may emerge from information-theoretical quantities in a gravitational dual \cite{Doi:2022iyj, Doi:2023zaf, Takayanagi:2025ula}. Moreover, given the strong interplay between quantum chaos and gravity \cite{Maldacena:2015waa}, it stands to reason that leveraging this new relation might allow for the construction of quantities capable of describing temporal correlations as a whole and providing clear signatures of chaos. However, while \cite{Doi:2023zaf} proved that the definition of a ``timelike entanglement entropy" via analytical continuation is related to the so-called pseudo-entropy \cite{Nakata:2020luh, Mollabashi:2020yie, Mollabashi:2021xsd, Caputa:2024gve}, a clear physical interpretation for quantum systems without gravity was still lacking. 

Fortunately, a recent proposal that provides a purely quantum understanding of timelike entanglement is that of a \textit{spacetime density matrix} — an operator whose trace pairing reproduces two-time Wightman functions \cite{Milekhin:2025ycm, Guo:2025dtq}. The spacetime density matrix, which we refer to as a \textit{space-time density kernel} for reasons that will become clear, generalises the standard reduced density matrix, to which it reduces upon taking a partial trace. It was also proven in \cite{Milekhin:2025ycm} that the trace of its integer powers (which are used to define the respective ``Renyi entropies") are equal to the ``Wick rotation" performed in \cite{Doi:2022iyj, Doi:2023zaf} of $\text{Tr}\rho_A^n$ for a local reduced density matrix $\rho_A$ as part of the replica trick. Thus, not only are the space-time density kernels objects which encode correlations between subsystems at different instances of time in the same way density matrices do for a single Cauchy slice, but their properties can also characterize causal relations in general, whether via timelike entanglement in gravity duals or through the kernels' general non-Hermitian nature representing the arrow of time, as taking their adjoint flips the time ordering of operators in a correlator \cite{Milekhin:2025ycm, Guo:2025dtq}. 

One point to note, however, is that the formalism developed so far is built upon two-point functions and its direct application to diagnoses of scrambling (a phenomenon intrinsically linked to four-point correlators, such as the OTOC) is not immediately clear. Building on these ideas, thus, we propose to lift the existing construction to higher-point correlations, so that OTOC and its finite-temperature regularisations become linear functionals of a single, tensorial kernel.

In this paper, we introduce a \textit{generalised spacetime density kernel} (GSDK), an operator kernel designed to encode multi-point, two-time correlation functions. We show that its algebraic structure encodes familiar chaos diagnostics, such as SFFs and OTOCs, as well as new bounds.  We first define a canonical two-leg operator $T_{AB}(t)$ acting on $A \otimes B$ such that for every pair of probes ($O_A, O_B$), $\mathrm{Tr}\left[T_{AB}(t), (O_A \otimes O_B)\right] = \mathrm{Tr}\left(\rho \;O_A O_B(t) \right)$.
This is the same object in \cite{Milekhin:2025ycm}, presented here in a channel-independent, operator-basis-friendly gauge; its adjoint flips the operator ordering. Next, we focus on the construction of a four-leg operator, $T^{(4)}(t)$, acting on four copies of the system's Hilbert space, which packages the full operator dependence of four-point correlators. This operator is defined such that the thermally-regularised OTOC is recovered by a simple trace pairing. 

We show a direct relationship between this GSDK and the spectral form factor. By performing probe averaging we map the two-leg kernel into the 2-point SFF, and $T^{(4)}(t)$ to powers of the analytically continued partition function, $|Z(\beta/4, t)|^4$. Thus, we find a link between GSDK and quantum chaos, as encoded in $T^{(4)}(t)$, and the universal spectral statistics of the system. We use this new framework to explore several bounds of quantum chaos. Because OTOCs are linear pairings with the GSDK, Hölder’s inequality immediately yields an upper bound in terms of Schatten norms, $|T^{(4)}(t)|_p$. By analysing the structure of the GSDK's coefficients in the context of the Eigenstate Thermalisation Hypothesis (ETH), we recover the universal bound on the Lyapunov exponent, $\lambda_L \le 2\pi/\beta$. Furthermore, we prove that the Schatten-$p$ norms of both GSDKs, the  $T^{(2)}(t)$ and $T^{(4)}(t)$, are independent of time for uniform thermal insertions and act as upper bounds on the two and four-point analytically continued partition functions up to some constant factors,  $|Z(\beta/2, t)|^2$ and  $|Z(\beta/4, t)|^4$ respectively.

This paper is organised as follows. In Section \ref{sec:def:GSDK}, we introduce the two-leg space-time density kernel derived from the two-point correlation function and establish our conventions. We also construct the four-leg generalised kernels for out-of-time-order correlators (OTOCs) and time-order correlators (TOCs), and demonstrate that their trace pairings reproduce the desired four-point functions for arbitrary probes. Subsequently, we outline the natural $2N$-leg extension of the GSDK construction, demonstrating how higher-point space-time kernels may be defined and how their pairings generate general $2N$-point correlation functions in a unified framework. In Section \ref{sec:sff}, we derive exact identities that link the simple probe averages of these kernels with various spectral form factors. In Section \ref{sec:bounds}, we establish bounds utilising properties of the GSDK. Specifically, in \ref{sec:Holder's bound}, we apply Hölder's inequality to derive upper bounds on four-point correlators through the norms of the GSDK. In \ref{sec:ETH bound} connects our formalism to the Eigenstate Thermalisation Hypothesis, illustrating how it constrains the growth of chaos. We conclude with a discussion in Section \ref{sec:discussion}. 
The appendices provide detailed proofs of the Haar averaging identities and our calculations pertaining to the time-independent Schatten norms of the GSDKs.

\section{Spacetime density kernel}\label{sec:def:GSDK}

\subsection{Spacetime kernel from two-point correlator}\label{sec:intro}

In this section, we express the spacetime density matrix which we refer to as spacetime density kernel using an operator basis for a generic structure of subsystem in quantum mechanical systems. We write all two-time correlators $\mathcal{L}_t(O_A, O_B) := \mathrm{Tr}\!\big(\rho\,O_A\,U_t^\dagger\,O_B\,U_t\big)$ with insertion on $A$ at time $0$ and on $B$ at time $t$ into a single kernel $T_{AB}(t)$ acting on $A \otimes B$. We want for every pair of probes $O_A, O_B$
\begin{equation}
\label{eq:target}
\mathcal{L}_t(O_A, O_B) = \mathrm{Tr}\!\big(T_{AB}(t)\,(O_A \otimes O_B)\big).
\end{equation}
We write the full Hilbert space as $S$. We pick subsystems $A \subset S$ and $B \subset S$ with complements $A^c, B^c$, so $S \simeq A \otimes A^c \simeq B \otimes B^c$. We choose orthonormal bases $\{|i\rangle_A\}$ on $A$ and $\{|k\rangle_B\}$ on $B$. The trace is defined on the state Hilbert space and is carried out over the local orthonormal bases, which will often be chosen as the energy eigenbasis. We define an operator basis as\footnote{Note that, this is the standard basis on the Hilbert space of operators, which is also known as the Hilbert-Schmidt space.}
\begin{equation}
E^A_{ij} := |i\rangle\!\langle j|_A, \qquad E^B_{kl} := |k\rangle\!\langle l|_B.
\end{equation}
Any operator on $A$ and respectively on $B$ is a linear combination of these. We embed these operators into the full space by acting trivially on the complements
\begin{equation}
\widehat{E}^A_{ij} := E^A_{ij} \otimes \mathbf{1}_{A^c}, \qquad \widehat{E}^B_{kl} := E^B_{kl} \otimes \mathbf{1}_{B^c}.
\end{equation}
We write the global state that can be either mixed or pure as $\rho$. We write Heisenberg evolution as $U_t=e^{-iHt}$, so $\Phi_t(X) := U_t^\dagger X U_t$ is the superoperator or channel acting on operators. Motivated by the definition used in \cite{Milekhin:2025ycm}, we would like $T_{AB}(t)$ to satisfy the identity,
\begin{equation}
\label{eq:two}
\mathrm{Tr}\!\big(T_{AB}(t)\,(O_A \otimes O_B)\big) = \mathrm{Tr}\!\big(\rho\, O_A \, U_t^\dagger O_B U_t\big).
\end{equation}
We first expand the probes in the corresponding basis
\begin{equation}\label{opt:def}
O_A=\sum_{i,j} O^{(A)}_{ij}\,E^A_{ij},\qquad O_B=\sum_{k,l} O^{(B)}_{kl}\,E^B_{kl}.
\end{equation}
Next, evaluate the RHS of \eqref{eq:two} using linearity and \eqref{opt:def} 
\begin{align}
\mathrm{Tr}\!\big(\rho\, O_A \, U_t^\dagger O_B U_t\big) &= \sum_{i,j,k,l} O^{(A)}_{ij}\,O^{(B)}_{kl}\; \mathrm{Tr}\!\big(\rho\,\widehat{E}^A_{ij}\,U_{t}^\dagger\,\widehat{E}^B_{kl}\,U_{t}\big)
= \sum_{i,j,k,l} O^{(A)}_{ij}\,O^{(B)}_{kl}\; C_{ij;kl}(t)\label{2pt:grf}
\end{align}
We tabulate the physical correlator on a basis by placing basis elements exactly where the real probes live in the correlator—this preserves operator ordering. We call $C_{ij;kl}(t)$ as the ``\textit{generalized response tensor}''. On $A$, $E_{ij}$ acts like a transition $|j\rangle \to |i\rangle$; on $B$, $E_{kl}$ probes $|l\rangle \to |k\rangle$ after the evolution $t$. So $C_{ij;kl}(t)$ evaluates in state $\rho$, the amplitude for $A:j \to i$ at $0$ and $B:l \to k$ at $t$. 

We turn the tensor $C_{ij;kl}(t)$ into an operator on $A\otimes B$ that reproduces the correlator by trace pairing. We write
\begin{equation}
\label{eq:one}
T_{AB}(t) = \sum_{i,j,k,l} C_{ij;kl}(t)\; E^A_{ji} \otimes E^B_{lk}. 
\end{equation}
The flipping of indices in $E_{ji}, E_{lk}$ follows from $\mathrm{Tr}(E_{ji}\,X)=X_{ij}$. So $\mathrm{Tr}(E^A_{ji}O_A)=O^{(A)}_{ij}$ and $\mathrm{Tr}(E^B_{lk}O_B)=O^{(B)}_{kl}$. For consistency check, we write the trace with $T_{AB}$ using \eqref{eq:one}:
\begin{align}
\mathrm{Tr}\!\big(T_{AB}(t)(O_A \otimes O_B)\big) &= \sum_{i,j,k,l} C_{ij;kl}(t)\; \mathrm{Tr}(E^A_{ji}O_A)\;\mathrm{Tr}(E^B_{lk}O_B) \nonumber\\
&= \sum_{i,j,k,l} C_{ij;kl}(t)\;O^{(A)}_{ij}\,O^{(B)}_{kl} \ .
\end{align}
The sums match \eqref{2pt:grf} term by term. Hence, this holds for all $O_A, O_B$. In this operator-independent canonical gauge, the ordinary adjoint satisfies
\begin{equation}
\mathrm{Tr}\!\big(T_{AB}(t)^\dagger\,(O_A \otimes O_B)\big) = \mathrm{Tr}\!\big(\rho\, O_B(t)\,O_A\big) \ ,
\end{equation}
 This identity can be shown explicitly by using the definitions \eqref{2pt:grf} and \eqref{eq:one},
\begin{align}
    T_{AB}(t)^\dagger
    =\sum_{i,j,k,l} C_{ij;kl}(t)^*\; E^A_{ij}\!\otimes E^B_{kl} \ , \label{eq:adjoint_T}
\end{align}
and evaluating $C_{ij;kl}(t)^*$ as $
    \operatorname{Tr}\!\Big(\big(\rho\,\widehat E^A_{ij}\,U_t^\dagger\,\widehat E^B_{kl}\,U_t\big)^\dagger\,\Big) =\operatorname{Tr}\!\big(\rho\,U_t^\dagger\,\widehat E^B_{lk}\,U_t\,\widehat E^A_{ji}\big)$ \ .
Then we obtain the following:
\begin{align}
   \operatorname{Tr}\!\big(T_{AB}(t)^\dagger\,(O_A\!\otimes\!O_B)\big)
    &=\sum_{i,j,k,l} C_{ij;kl}(t)^*\;\operatorname{Tr}(E^A_{ij}O_A)\;\operatorname{Tr}(E^B_{kl}O_B) =\mathrm{Tr}\!\big(\rho\, O_B(t)\,O_A\big) \ , \label{eq:contraction}
\end{align}
where we have used \eqref{opt:def}. Now, for a sanity check we take a partial trace over the subsystem $B$ in \eqref{eq:one}, using $\mathrm{Tr}_B(E^B_{lk})=\delta_{lk}$, and summing over $k$, this returns static reduced states,
\begin{align}
\mathrm{Tr}_B\,T_{AB}(t) &= \sum_{i,j,k} C_{ij;kk}(t)\,E^A_{ji} = \sum_{i,j} \mathrm{Tr}(\rho\,\widehat{E}^A_{ij})\,E^A_{ji} = \rho_A.
\end{align}
Also, tracing over the $A$ subsystem gives, $\mathrm{Tr}_A\,T_{AB}(t)=\rho_B(t)$.  In terms of dimensions, $T_{AB}(t)$ is a $d_A d_B \times d_A d_B$ matrix. As for special cases, if $A=B=S$, we recover the time-independent two-leg object.
\subsection{Generalised spacetime density kernel from higher point correlators}\label{sec:def GSDK2}

Motivated by the spacetime density kernel from the two point function, we define a generalized spacetime density kernel for four-point correlators. Higher-point correlators can be either time-ordered or out-of-time-ordered structures, depending on how operator insertions are arranged. The corresponding generalized spacetime density kernels differ in their coefficient structure, encoded in the generalized response tensor. We start with the out-of-time-order correlator with uniform thermal insertion $ \mathrm{Tr}\big[y\,V\,y\,W(t)\,y\,V\,y\,W(t)\big]$, 
with $V=V(0)$, $W(t):= U_t^\dagger W U_t$, $U_t=e^{-iHt}$, and $y^4=Z(\beta)^{-1}e^{-\beta H}$. We introduce four bookkeeping copies of the Hilbert space of $S$: label them $r=1,2,3,4$. On each copy, fix the matrix-unit basis
\begin{equation}
    E^{(r)}_{ij}:=|i\rangle\!\langle j| \quad (i,j=1,\dots,d_S) \ . \label{eq:basis_E}
\end{equation}
When we ``place'' a basis element into the physical trace on $S$, we write a hat:
\begin{equation}
    \widehat E^{(r)}_{ij}:=E_{ij} \quad\text{(as an operator on the physical }S\text{)} \ . \label{eq:hat_E}
\end{equation}
The superscript $(r)$ just remembers which leg it belongs to when we reconstruct $T$.
For matrix units and trace function of Kronecker product, we will repeatedly use the following identities. 
\begin{align} \label{eq:HS} &\mathrm{Tr}\!\big(E_{ji}\,X\big)=X_{ij},\,    \mathrm{Tr}\!\big(E_{ab}\,E_{cd}\big)=\delta_{bc}\,\delta_{ad} \ ,\\  &\mathrm{Tr}[(X_1\!\otimes\!\cdots\otimes\!X_4)(Y_1\!\otimes\!\cdots\!\otimes\!Y_4)] = \prod_{r=1}^4 \mathrm{Tr}(X_rY_r) \ . \label{trace:fun}
\end{align}
Here, $X$ and $Y$ are general self-adjoint operators.

For arbitrary probes $O_1,O_2,O_3,O_4$ on $S$ we define:
\begin{equation}
    \mathcal F_t[O_1,O_2,O_3,O_4]
    :=\mathrm{Tr}\!\Big(y\,O_1\,y\,U_t^\dagger O_2 U_t\,y\,O_3\,y\,U_t^\dagger O_4 U_t\Big) \ . \label{eq:otoc2}
\end{equation}
 This is the same operator ordering as the OTOC, but with $V,W,V,W$ replaced by a general operator, denoted by $O_r$. Our goal is to build a single operator $T^{(4)}(t)$ on $S^{\otimes4}$ such that
\begin{equation}
    \mathcal F_t[O_1,O_2,O_3,O_4]
    =\mathrm{Tr}\!\Big(T^{(4)}(t)\,(O_1\otimes O_2\otimes O_3\otimes O_4)\Big)
    \quad\text{for all }O_r \ . \label{eq:otoc3}
\end{equation}
We expand probes by writing each $O_r$ in the matrix-unit basis on its leg
\begin{equation}
    O_1=\sum_{i,j} O^{(1)}_{ij}\,E^{(1)}_{ij} \ ,\ 
    O_2=\sum_{k,l} O^{(2)}_{kl}\,E^{(2)}_{kl} \ ,\ 
    O_3=\sum_{m,n} O^{(3)}_{mn}\,E^{(3)}_{mn} \ ,\ 
    O_4=\sum_{p,q} O^{(4)}_{pq}\,E^{(4)}_{pq} \ . \label{opt:def2}
\end{equation}
Inserting \eqref{opt:def2} into \eqref{eq:otoc2} we rewrite $\mathcal F_t[O_1,O_2,O_3,O_4]$ as:
\begin{align}
    \sum_{i,j,k,l,m,n,p,q}
O^{(1)}_{ij}\,O^{(2)}_{kl}\,O^{(3)}_{mn}\,O^{(4)}_{pq} 
    \mathrm{Tr}\!\Big(
    y\,\widehat E^{(1)}_{ij}\,y\,
    U_t^\dagger\,\widehat E^{(2)}_{kl}\,U_t\,
    y\,\widehat E^{(3)}_{mn}\,y\,
    U_t^\dagger\,\widehat E^{(4)}_{pq}\,U_t
    \Big) \ . \label{eq:otoc4}
\end{align}
 Similarly to the two point density kernel, we define the following generalized response tensor for the four-point case:
\begin{equation}
    C_{ij;\,kl;\,mn;\,pq}(t)
    :=\mathrm{Tr}\!\Big(
    y\,\widehat E^{(1)}_{ij}\,y\,
    U_t^\dagger\,\widehat E^{(2)}_{kl}\,U_t\,
    y\,\widehat E^{(3)}_{mn}\,y\,
    U_t^\dagger\,\widehat E^{(4)}_{pq}\,U_t
    \Big) \ .
     \label{eq:star4}
\end{equation}
Then \eqref{eq:otoc4} becomes: 
\begin{equation}
    \mathcal F_t[O_1,O_2,O_3,O_4]
    =\sum_{i,j,k,l,m,n,p,q}
    C_{ij;\,kl;\,mn;\,pq}(t)\;
    O^{(1)}_{ij}\,O^{(2)}_{kl}\,O^{(3)}_{mn}\,O^{(4)}_{pq} \ . \label{final:otoc}
\end{equation}
Each element of the generalized response tensor is the amplitude/weight for doing the four elementary ``transitions''
($j\!\to\! i$) at the first slot, then ($l\!\to\! k$) at time $t$, then ($n\!\to\! m$), then ($q\!\to\! p$) at time $t$, with thermal factors $y$ seated between them exactly as in the OTOC. We use Hilbert--Schmidt duality \eqref{eq:HS} on each leg to read off the coefficients $O^{(r)}_{\cdot\cdot}$ by tracing against flipped matrix units. We define the generalized spacetime density kernel for four-point OTOC as:
\begin{equation}
    T^{(4)}(t)=\sum_{i,j,k,l,m,n,p,q}
    C_{ij;\,kl;\,mn;\,pq}(t)\;
    \Big(E^{(1)}_{ji}\otimes E^{(2)}_{lk}\otimes E^{(3)}_{nm}\otimes E^{(4)}_{qp}\Big) \ .
     \label{eq:1_4}
\end{equation}
Note that we flip the indices because for any $O$, $\mathrm{Tr}(E_{ji}O)=O_{ij}$ by \eqref{eq:HS}. The elements $E_{ji}$ ensures the trace pairing reads out the coefficient $O_{ij}$ with no extra transposes. Now we prove the target identity. Using \eqref{eq:HS} and \eqref{trace:fun} one can compute the trace in \eqref{eq:otoc3} to check that we indeed obtain $\mathcal F_t[O_1,O_2,O_3,O_4]$. 

Generally $T^{(4)}(t)$ is not hermitian. Changing $\{E^{(r)}_{ij}\}$ just changes the coordinates of both $T^{(4)}$ and the test operator $O_1\!\otimes\!\cdots\!\otimes\!O_4$ in a matched way; the identity \eqref{final:otoc} still holds for all $O_r$, so the underlying operator $T^{(4)}(t)$ is canonical up to that change of basis. We now check how the adjoint of the 4-point kernel behaves. By doing the adjoint operation, we get the following:
\begin{eqnarray}
T^{(4)}(t)^\dagger
    =\sum_{i,j,k,l,m,n,p,q}
    C_{ij;\,kl;\,mn;\,pq}(t)^{\!*}\ \big(E^{(1)}_{ij}\!\otimes E^{(2)}_{kl}\!\otimes E^{(3)}_{mn}\!\otimes E^{(4)}_{pq}\big) \ . 
\end{eqnarray}    
Now, we compute the trace-pairing with arbitrary probes $O_r=\sum O^{(r)}_{ab}E^{(r)}_{ab}$:
\begin{align}
    \mathrm{Tr}\!\left[T^{(4)}(t)^\dagger\,(O_1\!\otimes\!O_2\!\otimes\!O_3\!\otimes\!O_4)\right]
    &=\sum_{i,j,k,l,m,n,p,q} C_{ij;\,kl;\,mn;\,pq}(t)^{\!*}\, O^{(1)}_{ji}\,O^{(2)}_{lk}\,O^{(3)}_{nm}\,O^{(4)}_{qp} \ . \label{eq:adjoint_pairing}
\end{align}
By definition of the generalized response tensor, we obtain:
\begin{align}
    C_{ij;\,kl;\,mn;\,pq}(t)^{\!*}
    &=\mathrm{Tr}\!\Big[\,U_t^\dagger\,\widehat E^{(4)}_{qp}U_t\,y\,\widehat E^{(3)}_{nm}\,y\,U_t^\dagger\,\widehat E^{(2)}_{lk}U_t\,y\,\widehat E^{(1)}_{ji}\,y\Big] \ . \label{eq:C_conj}
\end{align}
Plugging this into the previous sum and using the expansions \eqref{opt:def2} the four-point function takes the form $\mathrm{Tr}\!\left[T^{(4)}(t)^\dagger\,(O_1\!\otimes\!O_2\!\otimes\!O_3\!\otimes\!O_4)\right]
    =\mathrm{Tr}\!\Big[\,U_t^\dagger O_4 U_t\,y\,O_3\,y\,U_t^\dagger O_2 U_t\,y\,O_1\,y\Big]$.
Finally, using $U_t^\dagger O_r U_t = O_r(t)$, we get the following:
\begin{equation}
    \ 
    \mathrm{Tr}\!\left[T^{(4)}(t)^\dagger\,(O_1\!\otimes\!O_2\!\otimes\!O_3\!\otimes\!O_4)\right]
    =\mathrm{Tr}\!\Big[\,y\,O_4(t)\,y\,O_3\,y\,O_2(t)\,y\,O_1\Big] \ . \label{eq:adjoint_result_heisenberg}
\end{equation}
Similarly to the adjoint of the two-point kernel, the adjoint flips the operator ordering along the contour: starting from the OTO order $1\to 2(t)\to 3\to 4(t)$, the adjoint produces $4(t)\to 3\to 2(t)\to 1$ --- i.e. reverse order with the same magnitude of time separation.

A similar GSDK can also be obtained for the time-ordered four-point correlator with equal-spacing thermal factors inserted $\mathrm{Tr}\!\big[\,y\,V\,y\,V\,y\,W(t)\,y\,W(t)\,\big]$.
We can define a single 4-leg operator $\widetilde{T}^{(4)}(t)$ on $S^{\otimes4}$ such that, for all probes,
\begin{equation}    \mathrm{Tr}\!\big[\,y\,V\,y\,V\,y\,W(t)\,y\,W(t)\,\big]
    =\mathrm{Tr}\!\Big[\widetilde{T}^{(4)}(t)\,\big(V\otimes V\otimes W\otimes W\big)\Big].
    \; \label{eq:Target_toc}
\end{equation}
For an arbitrary choice of $O_1,O_2,O_3,O_4$ on $S$, we set:
\begin{eqnarray}
\mathcal G_t[O_1,O_2,O_3,O_4]
    :=\mathrm{Tr}\!\Big(
    y\,O_1\,y\,O_2\,y\,
    U_t^\dagger O_3 U_t\,y\,
    U_t^\dagger O_4 U_t
    \Big) \ .
\end{eqnarray}    
This puts the first two insertions at the initial time with thermal insertions between them, then inserts the last two at time $t$, again separated by $y$'s. We proceed analogously to the earlier case by expanding each operator in the same operator basis \eqref{opt:def2}. Using the identities \eqref{eq:HS} and \eqref{trace:fun}, we then obtain the coefficients of the generalized spacetime density kernel corresponding to the four-point time-ordered correlator. 
\begin{equation}
    \;
    \widetilde{C}_{ij;\,kl;\,mn;\,pq}(t)
    :=\mathrm{Tr}\!\Big(
    y\,\widehat E^{(1)}_{ij}\,y\,\widehat E^{(2)}_{kl}\,y\;
    U_t^\dagger\,\widehat E^{(3)}_{mn}\,U_t\,y\;
    U_t^\dagger\,\widehat E^{(4)}_{pq}\,U_t
    \Big) \ .
    \; \label{eq:star_prime}
\end{equation}
So, we get:
\begin{equation}
    \mathcal G_t[O_1,O_2,O_3,O_4]
    =\sum_{i,j,k,l,m,n,p,q}
    \widetilde{C}_{ij;\,kl;\,mn;\,pq}(t)\;
    O^{(1)}_{ij}\,O^{(2)}_{kl}\,O^{(3)}_{mn}\,O^{(4)}_{pq} \ . \label{eq: TOC}
\end{equation}
From this we can reconstruct $\widetilde{T}^{(4)}(t)$ by the identity mentioned in \eqref{eq:HS},
\begin{equation}
   \;
    \widetilde{T}^{(4)}(t)=\sum_{i,j,k,l,m,n,p,q}
    \widetilde{C}_{ij;\,kl;\,mn;\,pq}(t)\;
    \Big(E^{(1)}_{ji}\otimes E^{(2)}_{lk}\otimes E^{(3)}_{nm}\otimes E^{(4)}_{qp}\Big) \ .
    \;\label{T4:for TO}
\end{equation}
One can easily check that taking the trace of $\widetilde{T}^{(4)}(t)$ with $O_{1}\otimes O_{2}\otimes O_{3}\otimes O_{4}$ reproduces \eqref{eq: TOC}.
The adjoint operator of $\widetilde{T}^{(4)}(t)$ flips the ordering of the operator insertions and produces the $\mathrm{Tr}\!\Big[\,y\,O_4(t)\,y\,O_3(t)\,y\,O_2\,y\,O_1\Big]$ time-ordered correlator.

This construction can be straightforwardly generalised to the $2N$-point functions. The special case of interest would be the $2N$-point OTOC. `For the $2N$-point generalisation, we consider $y^{2N}=Z(\beta)^{-1}e^{-\beta H}$ with each ``leg'' $r=1,\dots,2N$ fixing matrix units $E^{(r)}_{ij}=|i\rangle\langle j|$ and write $\widehat{E}^{(r)}_{ij}$ for the same operator embedded on the physical space. The  OTO pattern with equal-spacing generalises as: $\operatorname{Tr}\left[
y\,\widehat{E}^{(1)}_{i_1j_1}\,y\,U_t^{\dagger}\widehat{E}^{(2)}_{i_2j_2}U_t\,y\,\widehat{E}^{(3)}_{i_3j_3}\,\cdots\,y\,U_t^{\dagger}\widehat{E}^{(2N)}_{i_{2N}j_{2N}}U_t
\right]$.
For this correlator, we define a $2N$-leg generalised spacetime density kernel
\begin{equation}\label{eq:gen_spacetime_kernel}
T^{(2N)}(t)
=\sum_{{i_r,j_r}}
C_{i_1j_1;\dots;i_{2N}j_{2N}}(t)
\bigotimes_{r=1}^{2N} E^{(r)}_{j_ri_r} \ .
\end{equation}
 By the Hilbert–Schmidt duality, $\operatorname{Tr}(E^{(r)}_{j_ri_r}O_r)=O^{(r)}_{i_rj_r}$, this kernel reproduces the regulated ($2N$)-point functional
\begin{align}\label{eq:gen_functional}
\mathcal F_t[O_1,\dots,O_{2N}]
&=\operatorname{Tr}\left[T^{(2N)}(t)\,\bigotimes_{r=1}^{2N} O_r\right] \\
&=\operatorname{Tr}\left[y\,O_1\,y\,O_2(t)\,y\,O_3\,\cdots\,y\,O_{2N}(t)\right] \ , \nonumber
\end{align}
generalising the two- and four-point function. Through this construction, we highlight that a similar GSDK can be defined for any time-ordering of the $2N$-point function. 

\paragraph{Disjoint two-region OTOC AA–BB}
A seemingly straightforward generalisation can be made for OTOC of the form $ \mathrm{Tr}\big[y\,V\,y\,W(t)\,y\,V\,y\,W(t)\big]$ where we restrict the initial support of the operator $V$ and $W$ in subsystem A and B, respectively. A motivation for this choice is the fact that we would like to start with a sufficiently local operator while calculating the OTOC dynamics. But for our scenario the setup is much more general as these two subsystems can be spatially and temporally separated. Similar to the two point function with support of the operators in two different subsystems, we
assume $S=H_A\otimes H_B$, $\dim H_A=d_A$, $\dim H_B=d_B$, $L=d_A d_B$. Consider the disjoint OTOC of the form $\mathcal F_t[V_A,W_B]:=\Tr\!\Big(V_A(0)\,W_B(t)\,V_A(0)\,W_B(t)\Big)$, with $V_A$ acting non-trivially only on $H_A$ and $W_B$ only on $H_B$. Use the product matrix units
$E^A_{ij}=|i\rangle\langle j| \text{ on }H_A,\,
    E^B_{kl}=|k\rangle\langle l| \text{ on }H_B$.
We expand the operators with the support in the local subsystem in the respective basis and define the coefficients of the generalised response tensor
\begin{align}
    C_{ij;\,kl;\,mn;\,pq}^{A,B}(t)
    &:=\Tr\!\Big((E^A_{ij}\!\otimes I_B)\,U_t^\dagger(I_A\!\otimes E^B_{kl})U_t\,(E^A_{mn}\!\otimes I_B)\,U_t^\dagger(I_A\!\otimes E^B_{pq})U_t\Big),
\end{align}
and the kernel
\begin{equation}
    T^{(4)}_{A|B}(t)=\sum_{ij,kl,mn,pq}
    C_{ij;\,kl;\,mn;\,pq}^{A,B}(t)\,
    \big(E^{A,(1)}_{ji}\otimes E^{B,(2)}_{lk}\otimes E^{A,(3)}_{nm}\otimes E^{B,(4)}_{qp}\big). \label{eq:ss_otoc}
\end{equation}
While equation \eqref{eq:ss_otoc} is a direct generalization of the two-point function, the subsequent sections will demonstrate that this kernel yields richer temporal dynamics. Specifically, the GSDK defined in \eqref{eq:ss_otoc} is shown to possess a time-independent Schatten norm, in contrast to the constant Schatten norm of the kernel \eqref{eq:1_4}. These, and other main differences, will be discussed in the following two sections.

\section{Spectral form factor from generalised spacetime density kernel}\label{sec:sff}

Spectral statistics, long recognized as a universal signature of quantum chaos, encapsulate the correlations among a system's energy eigenvalues. The generalised spacetime density kernel (GSDK), as defined in the preceding section, is unique in that it incorporates both the density matrix and the spectral properties of the Hamiltonian. Consequently, this quantity provides a unified framework for studying the system's inherent spatial correlations and out-of-equilibrium dynamical features. In this section, we formally relate the spacetime density kernel to the Spectral Form Factor (SFF), beginning with the time evolution of the two-point correlation function under $H$ acting on an L-dimensional Hilbert space. For the analysis presented in this section, we emphasize that the operator basis elements $E_{ij}$ will be constructed using a global basis, specifically chosen to be the energy eigenbasis, unless explicitly stated otherwise.

To set up the probe average used in this section, we briefly recall the basic properties
of the normalized Haar measure \(d\mu(U)\) on \(U(L)\). The measure is left- and
right-invariant, and the associated conjugation twirl
\(\int_{U(L)} d\mu(U)\, U O U^\dagger\) projects any operator
\(O \in \mathrm{End}(\mathbb{C}^L)\) onto the identity component, proportional to
\(\mathbb{I}_L\). Here $\mathrm{End}(\mathbb{C}^L)$ denotes the algebra of linear operators acting on the $L$-dimensional Hilbert space $\mathbb{C}^L$. The nontrivial ingredient for our purposes is the averaging of the tensor product \(V \otimes V^\dagger\)
over Haar-random unitaries \(V \in U(L)\) yields the “swap average”
\begin{equation}\label{eq:second_moment}
\int_{U(L)} d\mu(V)\, V \otimes V^\dagger \;=\; \frac{1}{L}\,\mathrm{SWAP},
\end{equation}
where \(\mathrm{SWAP} = \sum_{i,j=1}^L |i,j\rangle\!\langle j,i|\) exchanges the two
tensor factors. A derivation, together with the corresponding one-point twirl
\(\int d\mu(U)\, U O U^\dagger = (\operatorname{Tr}O/L)\,\mathbb{I}_L\), is reviewed
in Appendix~\ref{appendix:A}.

 The utility of SWAP is twofold: first, the “swap trick”, $\operatorname{Tr}\!\big[(A\otimes B)\mathrm{SWAP}\big] = \operatorname{Tr}(AB)$
lets us contract indices in our kernels transparently; second, SWAP admits an expansion in any trace-orthonormal operator basis $\{E_\alpha\}$, $\mathrm{SWAP}=\tfrac{1}{L}\sum_\alpha E_\alpha\otimes E_\alpha$, and in particular for a single qubit $\mathrm{SWAP} = \frac{1}{2}\big(I\!\otimes\! I+\sigma_{x}\!\otimes\! \sigma_{x}+\sigma_{y}\!\otimes\! \sigma_{y}+\sigma_{z}\!\otimes\! \sigma_{z}\big)$.
Motivated by these Haar-average identities and the Pauli-basis decomposition, we define the family of swap (permutation) operators that “test” our multi-leg kernels—starting from the two-leg SWAP and extending to the 4-cycle tester used below—so that Haar averages collapse our expressions to simple contractions with permutation operators.
\subsection{2-point GSDK and SFF}

With the motivation from the uniform thermal insertions in OTOC, we work with the equal-spacing finite-temperature regularization for the two-point operator. With density matrix $\rho=y^4$ and $U_t=e^{-iHt}$ the two-leg kernel expressed in the global basis is
\begin{align}
\begin{split}
    T^{(2)}(t)=\sum_{i,j,k,l} C^{(\beta)}_{ij;kl}(t)\; E_{ji}\otimes E_{lk}, \qquad
    C_{ij;kl}^{(\beta)}(t)=\operatorname{Tr}\!\big(y\,E_{ij}\,y\,U_t^\dagger\,y\,E_{kl}\,y\,U_t\big). \label{eq:2.leg}
\end{split}
\end{align}
and it reproduces the two-point correlators: $\mathcal{L}_t(O_1,O_2)=\operatorname{Tr}\!\big(T^{(2)}(t)(O_1\otimes O_2)\big)$.
We employ \eqref{eq:second_moment} to relate the two-point spacetime density kernel to the $2^{\text{nd}}$ moment of the SFF. This is achieved by contracting of the two-point spacetime density kernel,
$T^{(2)}(t)$ with the SWAP operator, which effectively gives an average of the two-point correlator. 
\begin{align}
    \overline{F_2(t)}\;&:=\;\E_{O}\,\mathcal{L}_t(O,O^\dagger) \;\;=\;\frac{1}{L}\;\operatorname{Tr}\!\big[\,T^{(2)}(t)\,\mathrm{SWAP}\,\big]. \label{eq:avg}
\end{align}
Now insert $T^{(2)}(t)=\sum C^{(\beta)}_{ij;kl}(t)E_{ji}\!\otimes E_{lk}$ and use $\operatorname{Tr}[\mathrm{SWAP}(E_{ji}\!\otimes E_{lk})]=\operatorname{Tr}(E_{ji}E_{lk})=\delta_{il}\delta_{jk}$
\begin{equation}
   \overline{ F_2(t)}=\frac{1}{L}\sum_{i,j} C^{(\beta)}_{ij;ji}(t). \label{eq:collapse}
\end{equation}
\begin{figure}[t]
    \centering
     \begin{subfigure}[b]{0.32\textwidth}
     \centering
         \includegraphics[width=\textwidth]{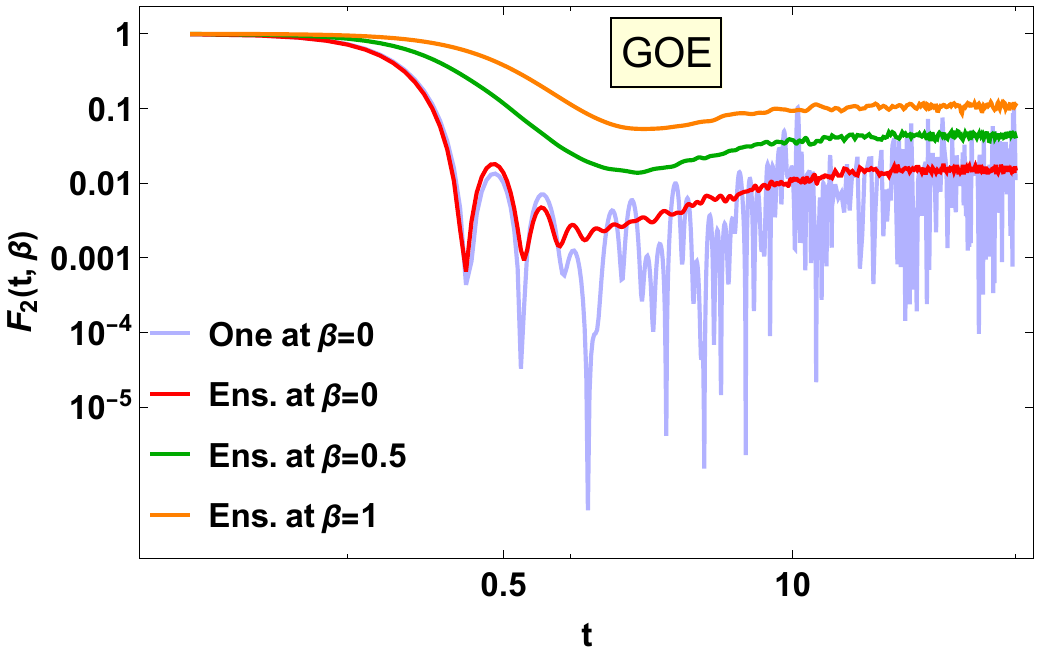}
         \label{fig:F2avggoe}
     \end{subfigure}
     \hfill
     \begin{subfigure}[b]{0.32\textwidth}
         \centering
         \includegraphics[width=\textwidth]{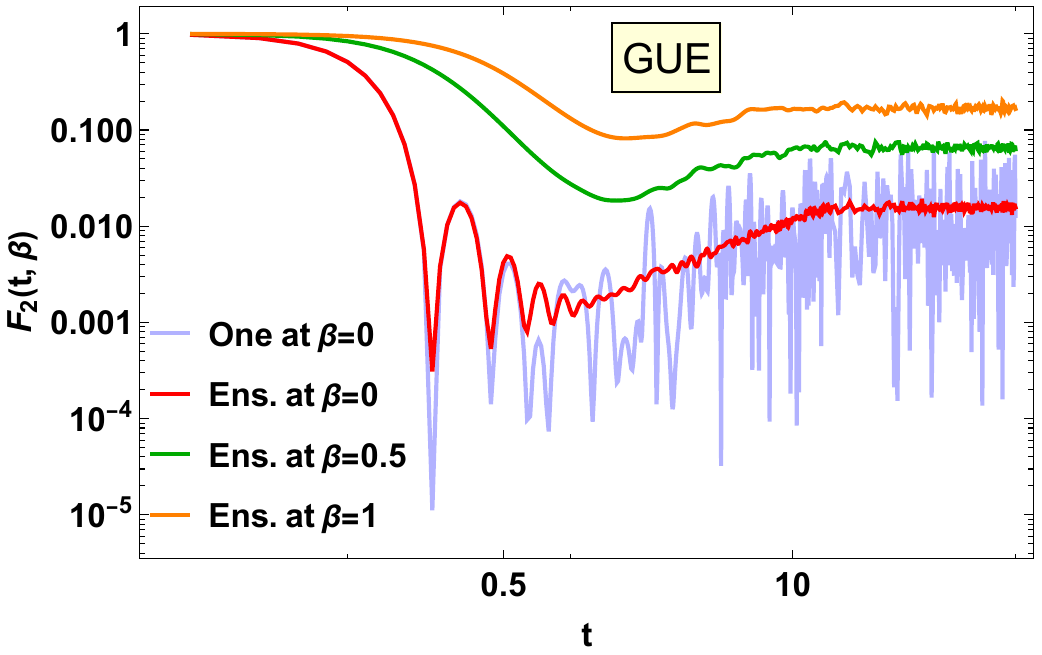}
         \label{fig:F2avggue}
     \end{subfigure}
     \hfill
     \begin{subfigure}[b]{0.32\textwidth}
         \centering
         \includegraphics[width=\textwidth]{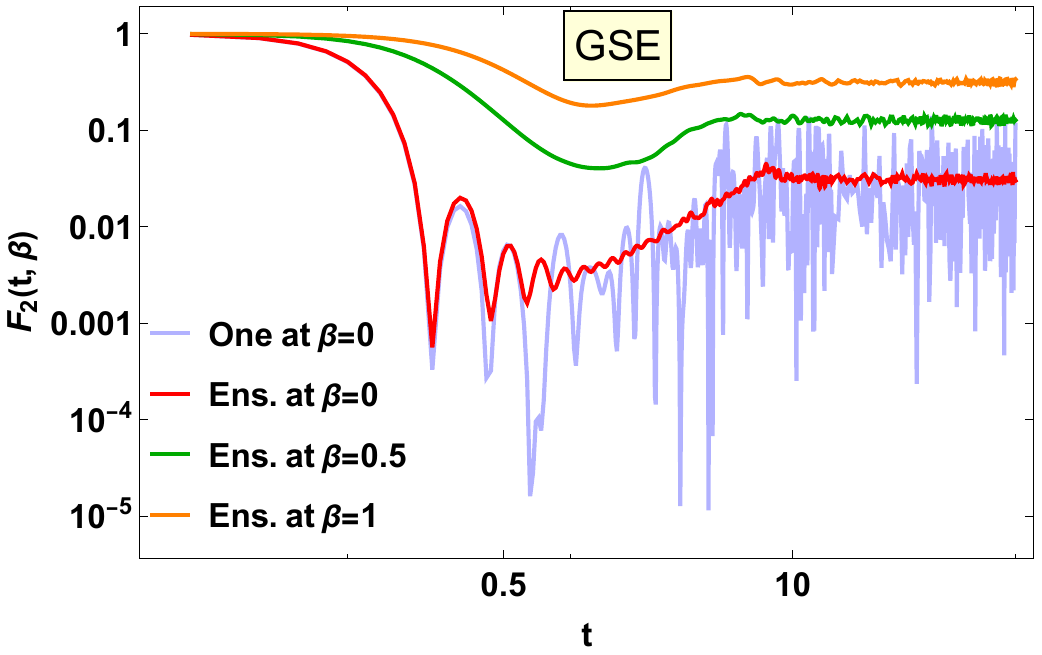}
         \label{fig:F2avggse}
     \end{subfigure}
        \caption{Normalised $F_2(t,\beta)$ as a function of time for Hamiltonian chosen from GOE, GUE and GSE distributions for $\beta=0,0.5,1$. The blue plot is for a single instance, and the red plot is for the average $\overline{F_2(t,\beta)}$ over 200 instances of random matrices.}
        \label{fig:complexity_werner_F2}
\end{figure}
\begin{figure}[hbtp]
    \centering
     \begin{subfigure}[b]{0.32\textwidth}
     \centering
         \includegraphics[width=\textwidth]{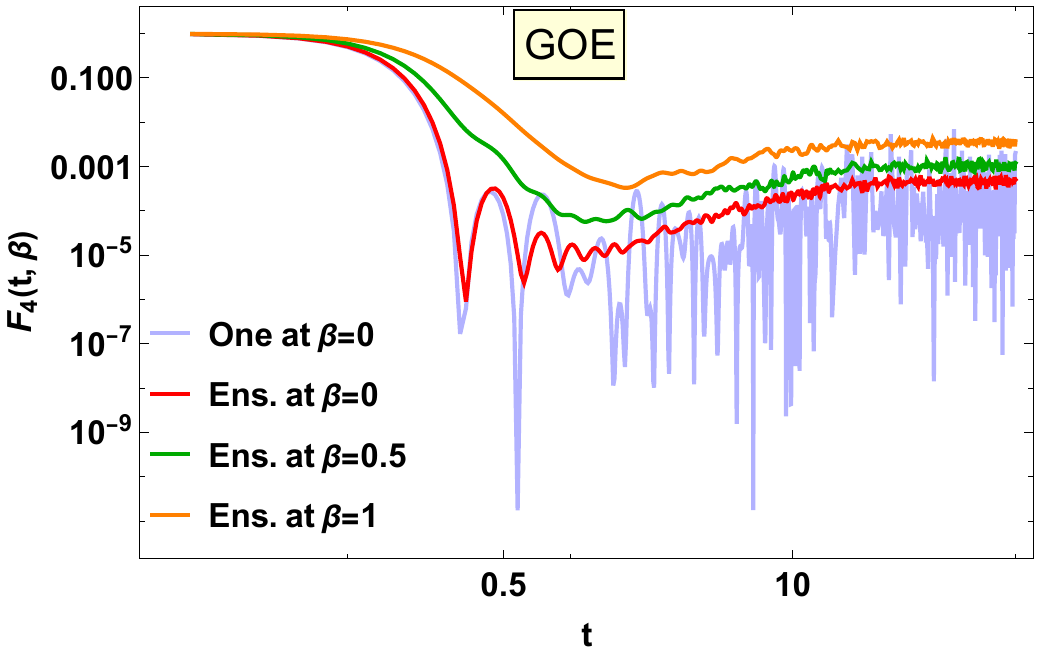}
         \label{fig:F4avggoe}
     \end{subfigure}
     \hfill
     \begin{subfigure}[b]{0.32\textwidth}
         \centering
         \includegraphics[width=\textwidth]{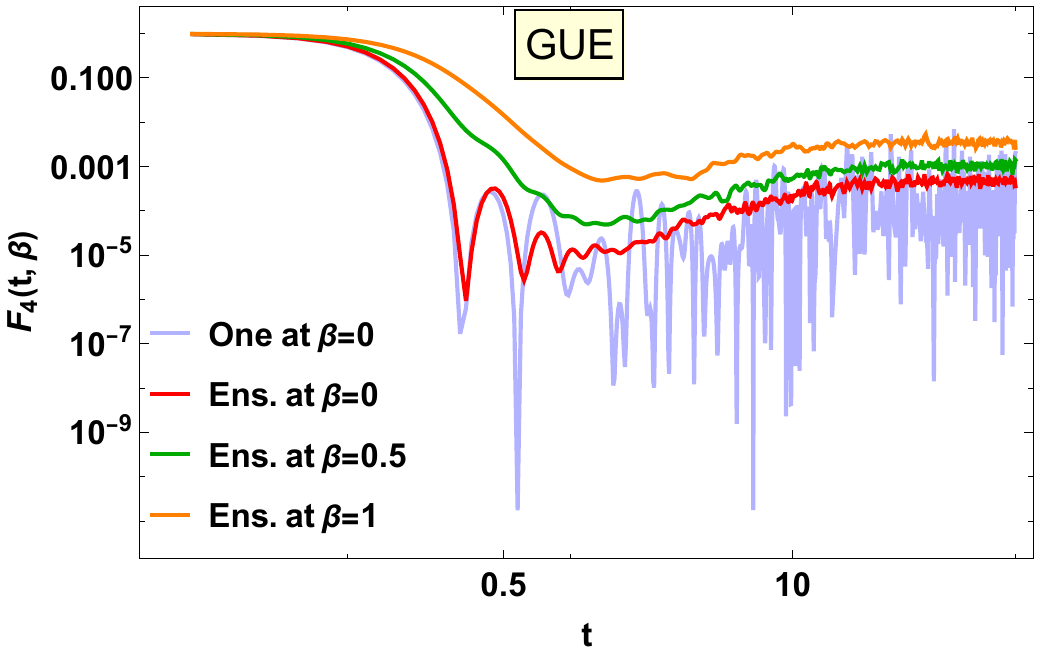}
         \label{fig:F4avggue}
     \end{subfigure}
     \hfill
     \begin{subfigure}[b]{0.32\textwidth}
         \centering
         \includegraphics[width=\textwidth]{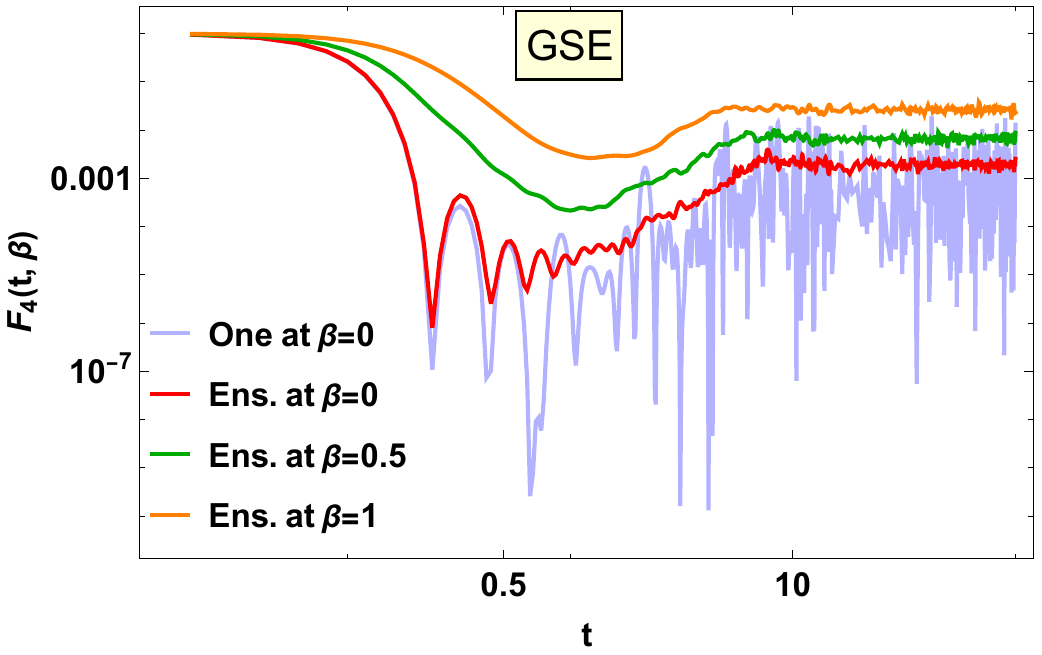}
         \label{fig:F4avggse}
     \end{subfigure}
        \caption{Normalised $F_4(t,\beta)$ as a function of time for Hamiltonian chosen from GOE, GUE and GSE distributions for $\beta=0,0.5,1$. The blue plot is for a single instance, and the red plot is for the average $\overline{F_4(t,\beta)}$ over 200 instances of random matrices.}
        \label{fig:complexity_werner_F4}
\end{figure}
In the equal-spacing finite-temperature regularization, $C^{(\beta)}_{ij;ji}(t)=\Tr \left(y^{2}E_{ij}y^{2}U^{\dagger}_{t}E_{ji}U_{t}\right) $.  As the global basis, we choose the Hamiltonian eigenbasis and we evaluate the conjugation of a matrix unit
$U_t^\dagger E_{ji} U_t
    = \Big(e^{i(E_j-E_i)t}\,E_{ji}\Big)$ and using $y^2 E_{ij} y^2=y^{2}_{i}y^{2}_{j}E_{ij}$, we get
\begin{equation}
    C^{(\beta)}_{ij;ji}(t)
    = y_i^2 y_j^2 e^{i(E_j-E_i)t}\;\operatorname{Tr}(E_{ij}E_{ji})=y_i^2 y_j^2 e^{i(E_j-E_i)t} .\label{eq:C_derivation}
\end{equation}
Plugging \eqref{eq:C_derivation} into \eqref{eq:collapse} gives the finite-temperature SFF 
\begin{equation}
   \quad \overline{F_2(t)}\;=\;\frac{1}{L}\,\big|\operatorname{Tr}(y^{2}U_t)\big|^{2}
    \;\equiv\;\frac{1}{L}\frac{\left|Z(\beta/2,t)\right|^2}{\left|Z(\beta)\right|} \ ,
\end{equation}
where the partition function is $Z(\beta)={\rm Tr}[e^{-\beta H }]$.
As a quick check, at infinite temperature ($\beta=0$) we have $y^{2}=\mathbb{I}$ so
\begin{equation}
    \overline{F_2(t)}=\frac{1}{L}\,|\operatorname{Tr}(U_t)|^{2},
\end{equation}
which is the standard spectral form factor at infinite temperature up to the expected $1/L$ from the single second-moment average. 
Instead of working with the equal-spacing, finite-temperature regularized two-point function, we may retain the entire thermal weight $\rho=y^{4}$ sitting entirely in front, exactly as in \eqref{eq:two}
\begin{equation}
    C^{(\beta)}_{ij;kl}(t)=\operatorname{Tr}\!\big(y^{4}E_{ij}U_t^\dagger E_{kl}U_t\big),
\end{equation}
then the same computation gives:
\begin{align}
    \overline{F_2}^{\text{(canon)}}(t)&=\frac{1}{L}\sum_{i,j}\operatorname{Tr}\!\big(y^{4}E_{ij}U_t^\dagger E_{ji}U_t\big) =\frac{1}{L}\,\operatorname{Tr}(U_t)^{*}\;\operatorname{Tr}(y^{4}U_t) \ . \label{eq:F2haar}
\end{align}
This is a mixed-temperature SFF $\overline{F_2}^{\text{(canon)}}(t)=\frac{1}{L \;Z(\beta)}\,Z(0,t)^{*}\,Z(\beta,t)$. Note that, this is not a manifestly real-valued quantity. It is basically the geometric mean of the SFF at $\beta=0$ and $\beta\not=0$. At this point, it is worth emphasizing again that the equal-spacing regularized correlator, upon a Haar-averaging, yields the SFF. Other regularizations are not natural from this point of view.  

\subsection{4-point GSDK, OTOC, TOC and SFF}

Now following the results of the two-leg Kernel, we study the relation between $4$-point kernel and its possible reduction to the SFF. For that, we first write the four-leg tester  as
\begin{equation}
    \Omega_{(1234)}\ = \sum_{a,b,c,d=1}^L
    \;E_{ab}\ \otimes\ E_{bc}\ \otimes\ E_{cd}\ \otimes\ E_{da}\label{eq:tester_def}
\end{equation}
This is a 4-leg extension of the SWAP operator.\footnote{This is actually a four-cycle permutation operator. Given the SWAP operator, it is straightforward to generalize it to permutation operators with an arbitrary number of levels. See appendix \ref{appendix:D} for a brief discussion.} When acting on a basis state $(|i\rangle\otimes|j\rangle\otimes|k\rangle\otimes|l\rangle)$ in  $L^{4}$-dimensional Hilbert space, the tester implements a cyclic permutation, yielding $(|l\rangle\otimes|i\rangle\otimes|j\rangle\otimes|k\rangle)$. Before evaluating the trace  of $\Omega_{(1234)}$ against the four-point kernel \eqref{eq:1_4}, we first express the latter in the global matrix unit basis: 
\begin{equation}
   T^{(4)}(t)
    =\sum_{i,j,k,l,m,n,p,q}
    C^{(\beta)}_{ij;\,kl;\,mn;\,pq}(t)\;
    \big(E_{ji}\!\otimes E_{lk}\!\otimes E_{nm}\!\otimes E_{qp}\big) \ . 
\end{equation}
The associated generalized response tensor is:
\begin{equation}
   C^{(\beta)}_{ij;\,kl;\,mn;\,pq}(t)
    =\operatorname{Tr}\!\Big(yE_{ij}y\;U_t^\dagger E_{kl}U_t\; yE_{mn}y\; U_t^\dagger E_{pq}U_t\Big). \label{eq:C_def} 
\end{equation}
Now we calculate the contraction $\overline {F_{4}(t)}=\frac{1}{L}\operatorname{Tr}\!\big[T^{(4)}(t)\,\Omega_{(1234)}\big]$, by
inserting the expansions and tracing the spacetime kernel $T^{(4)}(t)$ with the four-leg tester $\Omega_{(1234)}$ yields
\begin{equation}
      \overline{F_4(t)}=\frac{1}{L}\sum_{i,j,l,m}
    C^{(\beta)}_{ij;\,j\,l;\,l\,m;\,m\,i}(t) \ . \label{eq:star}
\end{equation}
Expressing the generalised response tensor in the energy eigenbasis and
organizing the average four-point OTOC in terms of energy differences and Boltzmann factors, $y^4=\rho_\beta=\dfrac{e^{-\beta H}}{\operatorname{Tr}(e^{-\beta H})}$  manifest a structure that is ultimately governed by the spectral form factor.
\begin{equation}
    \overline{F_4(t)}\;=\;\frac{1}{L}\big|\operatorname{Tr}\!\big(\rho_\beta^{1/4}\,U_t\big)\big|^{4} \equiv\;\frac{1}{L}\frac{\left|Z(\beta/4,t)\right|^4}{\left|Z(\beta)\right|} \ . \label{eq:Fbar_normalized}
\end{equation}
At $\beta=0$, it is the standard SFF upto a factor of $\frac{1}{L}$. The structure exhibited above extends straightforwardly to $2N$-point OTOCs and, correspondingly, to the associated $2N$-point SFF. For a given operator ordering specified in \eqref{eq:gen_functional}, 

\begin{equation} \label{eq:2N SFF}
    \overline{F_{(2N)}(t)}=\frac{1}{L}\operatorname{Tr}\left(T^{(2N)}(t)\Omega_{(1,...,2N)}\right)=\frac{1}{L}\frac{\left|Z(\beta/2N,t)\right|^{2N}}{\left|Z(\beta)\right|}.
\end{equation}

Several comments are in order: First, it is evident that performing a Haar-averaging over a $2N$-point function directly yields the $(2N)$-moment of the SFF. These moments, also known as the higher point SFF, capture finer details of the level-correlations, compared to the standard SFF. As demonstrated in \cite{Legramandi:2024ljn}, by analyzing these higher moments, it is possible to distinguish between a truly RMT-class system from the one that produces an RMT-like linear ramp in SFF.\footnote{An explicit example of such a system is discussed in \cite{Das:2023yfj}, where a logarithmic deterministic spectrum gives rise to a linear ramp in the SFF. Clearly, the information of all moments of the SFF is equivalent to the full knowledge of the level-spacing-distribution of the given spectrum.} Secondly, it is interesting to note that for the Haar-avergaed $(2N)$-point correlation function, the $(2N)$-moment of the SFF is evaluated at an effective temperature $\beta_{\rm eff}=\beta/(2N)$. Evidently, in the limit $N\gg 1$, $\beta_{\rm eff}\ll 1$. Thus, not only does the higher point function, after Haar-averaging, produce a higher moment of the SFF, it also increases the effective temperature by a factor of $N$. This enhancement is correlated with a better emergence of the dynamical imprints of the underlying Hamiltonian, especially features like the dip-ramp-plateau. The higher point functions, therefore, are naturally capable of enhancing the chaotic features of the corresponding dynamics. Finally, also note the contrast between equation (\ref{eq:2N SFF}) and (\ref{eq:F2haar}): While the former is manifestly real-valued, the latter is not. Note that this difference stems from the difference in regularizing schemes of the underlying correlation function. Equation (\ref{eq:2N SFF}) is evaluated with an evenly-spaced factor of the (appropriate power of the) density matrix between the operators. On the other hand, (\ref{eq:F2haar}) has an uneven spacing of the density matrix between the operators. The result in (\ref{eq:2N SFF}) clearly demonstrates that the equal-spacing scheme contains a cleaner dynamical imprint of the system and is more natural in this context.\footnote{See, {\it e.g.}~\cite{Romero-Bermudez:2019vej} where the dependence of an OTOC with respect to the shape of the Schwinger-Keldysh contour has been explicitly demonstrated. Furthermore, it has been argued that the equal-spacing OTOC produces the physical answer which is consistent with the notion of the maximal chaos, when applicable. This is consistent with our observation.}


A similar analysis can be done for the GSDK corresponding to the 
four-point time ordered correlator (as defined in equation \eqref{T4:for TO}), using the same tester that has been introduced in equation \eqref{eq:tester_def}. The Haar-averaged four-point time-ordered correlator \eqref{eq: TOC} reduces to: 
\begin{equation}\label{eq:4pt TO SFF}
    \overline{G^{(4)}(t)}=\frac{1}{L}\;\frac{\big[Z(\beta/4)\big]^2}{Z(\beta)}\;
    \Big|Z(\beta/4,t)\Big|^2 \ .
\end{equation}
The computation shows that tracing the GSDK with the tester opeartor for the four-point time-ordered correlator reproduces the second-moment of SFF, evaluated at an effective inverse temperature $\beta/4$. More generally, performing the same analysis on a $2N$-point time-ordered correlator produces the $2^{\rm nd}$-moment of SFF at an effective temperature $\beta/2N$. In this sense, time-ordered correlators cannot probe physics beyond the $2^{\rm nd}$-moment, reflecting the limitation of time-ordered correlators even after being subjected to a Haar-averaging. On the other hand, out-of-time-ordered correlators provide an access to all order moments of the SFF. 
\subsection{Trace Relations for Temporally Separated GSDKs}

Let us now discuss trace properties of the GSDK and its two point functions for different density matrices in the global energy eigenbasis. Let us begin with the thermal density matrix $\rho_{\beta}=\frac{e^{-\beta H}}{Z(\beta)}$. This is a function of the Hamiltonian, it trivially commutes with $H$. Consequently, its time evolution is stationary, $\rho_{\beta}(t)=\rho_{\beta}(0)$ and the overlap, $\operatorname{Tr}\left(\rho_{\beta}(t)\rho_{\beta}(t')\right)$ is manifestly time independent. In the energy eigenbasis, this trace simply multiplies the Boltzmann weights $p_{n}=e^{-\beta E_{n}}/Z(\beta)$, yielding the familiar purity of the Gibbs state, $Z(2\beta)/Z(\beta)^{2}$. In contrast, for an initial thermofield double (TFD) state, given by 
\begin{equation}
    |\psi_{\beta}\rangle=\frac{1}{\sqrt{Z(\beta)}}\sum_{n}e^{-\frac{\beta E_{n}}{2}|n,n\rangle} \ ,
\end{equation}
the situation is dynamical. The corresponding density matrix: 
\begin{equation}
\rho_{\beta}=|\psi_{\beta}\rangle\langle\psi_{\beta}|=\frac{1}{Z(\beta)}\sum_{n,m}e^{-\frac{\beta (E_{n}+E_{m})}{2}} |n,n\rangle\langle m,m| \ ,
\end{equation}
evolves under $(H_{L}+H_{R})/2$, where $H_{L,H}$ act independently on the left and right copies of the Hamiltonian. The evolution takes the form:
\begin{equation}
    \rho_{\beta}(t)=e^{-i\frac{(H_{L}+H_{R}) t}{2}} \rho_{\beta} \;e^{i\frac{(H_{L}+H_{R}) t}{2}}=\frac{1}{Z(\beta)}\sum_{n,m}e^{-i (E_{n}-E_{m})t}e^{-\frac{\beta (E_{n}+E_{m})}{2}} |n,n\rangle\langle m,m| \ . 
\end{equation}
 The autocorrelation function of the density matrix, $\operatorname{Tr}(\rho_{\beta}(t)\rho_{\beta})$, of the TFD state is the spectral form factor (SFF) of the TFD $
    \operatorname{Tr}(\rho_{\beta}(t)\rho_{\beta})=\frac{|Z(\beta, t)|^2}{Z(\beta)}
$.
Motivated by this observation, we now carry out an analogous analysis for the GSDK. As in the preceding section, we work in the matrix unit basis $E_{ij}$, constructed with respect to a global Hilbert-space basis that is constructed with the energy eigenbasis. We begin by considering the two-point function of GSDK  written in the global basis \eqref{eq:2.leg}:
\begin{equation}\label{eq:trT2}
    \operatorname{Tr}\left(T^{(2)}(t)T^{(2)}(t')\right)=\sum_{i,j,k,l}C^{(\beta)}_{ij;kl}(t)C^{(\beta)}_{ji;lk}(t') \ , 
\end{equation}
where $C^{(\beta)}_{ij;kl}(t)$ is defined in \eqref{eq:2.leg}. Using the relations
$y^2\;E_{ij}\;y^2=y^2_{i}\,y^2_{j}\,E_{ij},$ and $U^{\dagger}_{t}E_{kl}U_{t}=e^{i(E_{k}-E_{l})t} E_{kl}$ and similarly for $t'$, a straightforward evaluation yields:
\begin{equation}
    \operatorname{Tr}\left(T^{(2)}(t)T^{(2)}(t')\right)=\frac{|Z(\beta,(t-t'))|^2}{Z(\beta)^2} \ .
\end{equation}
A similar computation for the out-of-time-ordered 4-pt GSDK yields: 
$
   \operatorname{Tr}\left(T^{(4)}(t)T^{(4)}(t')\right)=\frac{|Z(\beta,2(t-t'))|^2}{Z(\beta)^2}
$.
In contrast, $\operatorname{Tr}\left(\widetilde{T}^{(4)}(t)\widetilde{T}^{(4)}(t')\right)=1$, while $\operatorname{Tr}\left(\widetilde{T}^{(6)}(t)\widetilde{T}^{(6)}(t')\right)=\frac{|Z(\beta,(t-t'))|^2}{Z(\beta)^2}$.
It can be checked in a straightforward way that a two-point function of a GSDK produces:
\begin{align}\label{eq:trace_relations}
    \operatorname{Tr}\left(T^{(2n)}(t)T^{(2n)}(t')\right) &= \frac{|Z(\beta,n(t-t'))|^2}{Z(\beta)^2} \ , \\ \operatorname{Tr}\left(\widetilde{T}^{(2n)}(t)\widetilde{T}^{(2n)}(t')\right) &= 
\begin{cases}
1, & n\ \text{even},\\
\frac{|Z(\beta,t-t')|^2}{Z(\beta)^2}, & n\ \text{odd} \ .
\end{cases}
\end{align}
The scaling in time $t \to n t$ does not produce any new physical effect in the corresponding SFF, however, it can scale various time-scales accordingly. For example, consider the example of an one-dimensional simple harmonic oscillator, for which the SFF is given by
\begin{eqnarray}
    |Z(\beta, t)|^2 = \frac{e^{-\beta \omega}}{1+e^{-\beta\omega} - 2 e^{-\beta\omega} \cos(\omega t)} \ . 
\end{eqnarray}
The corresponding revivals occur at the minimum of the denominator, which corresponds to $\cos(\omega t) = 1$ and, therefore, $t_{\rm rev}= (2\pi)/\omega$. If we scale up $t\to n t$, then the revival-time is accordingly scaled down: $t_{\rm rev} = (2\pi)/(n\omega)$.

Consider a rational CFT now. For such cases, the partition function can be expressed as a finite sum over characters:
\begin{eqnarray}
    Z(\tau, {\bar \tau})=\sum_{i,j} M_{ij} \chi_i(\tau) \bar{\chi}_j({\bar \tau}) \ , \quad \chi_i = {\rm Tr}_{\mathcal V_i} q^{h-\frac{c}{24}} \ , \quad q = e^{2\pi i \tau} \ . 
\end{eqnarray}
To evaluate the SFF, we set $\tau = t/(2\pi) + i \beta/(2\pi)$. Under $t\to (n t)$, the characters transform as: $\chi(\tau) \to e^{-i (n-1) t c/24} \chi(\tau)$. However, the final effect on the SFF is to again scale the corresponding physical time-scales that are obtained from the SFF. For holographic and/or chaotic CFTs, a similar scaling of {\it e.g.}~dip, ramp and plateau time-scales can be explicitly obtained.


\section{Bounds from the spacetime kernel}\label{sec:bounds}

When discussing subsystem correlations along a single time slice, one of the key properties of the ``ordinary" von Neumann entropy (which, for globally pure states is an entanglement measure) is the fact that one of its specific measures, the mutual information serves as a bound on \textbf{all} correlations between bounded operators acting on the subsystems. More precisely, given two spacelike subsystems $\text{A}$ and $\text{B}$, \textit{any} choice of observables $O_{\text{A}}$, $O_{\text{B}}$ acting exclusively on each and a global state $\rho$, the mutual information $I(\text{A}:\text{B})$ in $\rho$, which may be defined both for tensor product Hilbert spaces and more general quantum systems \cite{Nielsen:2012yss, Araki:1976zv}, satisfies the inequality \cite{ohyapetz93, Wolf:2007tdq} 

\begin{equation}
\label{bound:spacelike_ent}
    I(\text{A}:\text{B}) \geq \frac{1}{2}\frac{\langle O_{\text{A}}O_{\text{B}}\rangle_c^2}{|O_{\text{A}}|_\infty^2|O_{\text{B}}|_\infty^2},
\end{equation}
where $\langle O_{\text{A}}O_{\text{B}}\rangle_c := \langle O_{\text{A}}O_{\text{B}}\rangle-\langle O_{\text{A}}\rangle\langle O_{\text{B}}\rangle$, denoting the connected, or truncated, correlation function of the observables in the state $\rho$ and $|O|_\infty$ their operator (supremum) norm.

Thus, even before introducing the idea of the entropy and mutual information as measures of how many Bell pairs may be distilled from a given pure state \cite{PhysRevA.51.2738, Nielsen:2012yss}, these quantities already have a clear physical meaning in terms of bounding all possible correlations between subsystems. The work in \cite{Milekhin:2025ycm} provided a similar interpretation for quantities which may be derived from their kernel $T_{AB}(t)$, which we shall reproduce and generalize in this Section. In doing so, it will be made clear how the generalized spacetime kernels are able to quantify, via bounds, the overall temporal correlations in quantum systems and provide a new interesting window towards the study of chaos from this perspective.

\subsection{Bounds using H\"older's inequality}\label{sec:Holder's bound}

As a starting point to derive these bounds on all timelike correlations and connect to the study in \cite{Milekhin:2025ycm}, it is important to notice that a GSDK $T^{(n)}(t_1,...,t_n)$ associated to a quantum state $\rho$ is mathematically given by a \textit{linear functional} of the form
\begin{equation}
    \begin{split}
        T^{(n)}&: \mathcal{A}_{S_1}\otimes...\otimes\mathcal{A}_{S_n} \to \mathbb{C}\\
        &T^{(n)}(O_{S_1}...O_{S_n}) = \text{Tr}(\rho O_{S_1}(t_1)...O_{S_n}(t_n)),
    \end{split}
\end{equation}
where $\mathcal{A}_{S_i}$ is the algebra of bounded observables acting on the subsystem $S_i$ and where the choice of $\rho$, the different time slices $\{t_1,...,t_n\}$ and whether or not to insert the factors of $y=\rho_\beta^{\frac{1}{n}}$ naturally define different spacetime kernels, as given by Eqs. (\ref{eq:otoc3}), (\ref{eq:Target_toc}) and (\ref{eq:gen_functional}). Regardless, the general structure of a linear functional acting on the tensor product of the operator algebras of subsystems remains.

Such an abstract characterization of the GSDKs allows for them to be defined both in lattice systems and QFTs, as the theory of operator algebras \cite{takesaki1, takesaki2} may be used in order to define meaningful quantities even in the continuum \cite{Hollands:2017dov}. However, since these kernels are new objects whose properties have not yet been fully explored, for the remainder of this work we will, for simplicity, continue to restrict ourselves to either finite-dimensional systems or infinite lattices with finite-dimensional Hilbert spaces per site, as we have been doing implicitly so far.

Now, as a generic $T^{(n)}(t_1,...,t_n)$ is a linear functional, and in this work a \textit{bounded} one, it satisfies the so-called H\"older's inequality \cite{lang}, which states that, given a general bounded operator $O \in \mathcal{A}_{S_1}\otimes...\otimes\mathcal{A}_{S_n}$, then

\begin{equation} 
\left|\operatorname{Tr}\left[T^{(n)}(t_1,...,t_n)O\right]\right|\le |T^{(n)}(t_1,...,t_n)|_p|O|_q,\qquad \frac{1}{p}+\frac{1}{q}=1, 
\end{equation}\label{eq:gsdk_bound1}
where $|O|_p:= \left(\text{Tr}\left[\left(O^\dagger O\right)^\frac{p}{2}\right]\right)^\frac{1}{p}$ is the Schatten $p$-norm, when it is well-defined (for many-body systems this may require restrictions on the choice of subsystems $S_i$, depending on the global state $\rho$ and value of $p$).

Given that the functional defined by $T^{(n)}(t_1,...,t_n)$ returns, by construction, the correlator of the chosen operators acting on subsystems, Eq. (\ref{eq:gsdk_bound1}) is, for $n=2$, exactly the inequality shown in \cite{Milekhin:2025ycm} and the timelike generalization of Eq. (\ref{bound:spacelike_ent}).

Looking now at higher values of $n$, we may use the multiplicativity of Schatten norms over tensor products $|O_A\otimes O_B|_p = |O_A|_p|O_B|_p$ to write $O = \otimes_{i=1}^nO_{S_i}$ and
\begin{equation}
    |T^{(n)}(t_1,...,t_n)|_p\prod_{i=1}^n|O_{S_i}|_q \geq \left|\operatorname{Tr}\left[T^{(n)}(t_1,...,t_n)O\right]\right|,\qquad \frac{1}{p}+\frac{1}{q}=1.
\end{equation}\label{eq:gsdk_bound2}

Thus, a better understanding of the time-dependence of $|T^{(n)}(t_1,...,t_n)|_p$ for various quantum systems may shed light on the nature of quantum chaos, as they place limits on the growth and behaviour of time correlations. In particular, applying the above inequality to the kernel $T^{(4)}(t)$ defined in Eqs. (\ref{eq:otoc3}) and (\ref{eq:1_4}) leads to
\begin{equation} \label{eq:14} 
\begin{aligned}
|F_t[O_1,O_2,O_3,O_4]| &= \big|\operatorname{Tr}\!\big[T^{(4)}(t),(O_1 \otimes O_2 \otimes O_3 \otimes O_4)\big]\big| \\
&\le |T^{(4)}(t)|_p |O_1 \otimes O_2 \otimes O_3 \otimes O_4|_q \\
&=  |T^{(4)}(t)|_p \prod_{r=1}^4|O_r|_q,
\end{aligned}
\end{equation}
showing that the norm of the appropriate spacetime kernel gives us direct access to properties of \textit{all} out-of-time-order correlators. Choosing, in particular, $(O_1,O_2,O_3,O_4)=(W,V,V,W)$, results in
\begin{equation} \label{eq:21}
|T^{(4)}(t)|_p |V|_q^2 |W|_q^2 \geq |\operatorname{Tr}[T^{(4)}(t)(V \otimes W \otimes V \otimes W)]|,
\end{equation}
a bound on the ``traditional" form of the OTOC \cite{Maldacena:2015waa}.

It is worth pointing out that other quantities derived from $T^{(n)}(t_1,...,t_n)$ provide bounds on correlators which may be of interest. The authors in \cite{Milekhin:2025ycm} showed that
\begin{equation} \label{eq:16}
|\langle[O_A(0),O_B(t)]\rangle|=\big|\operatorname{Tr}\big[(T_{AB}-T_{AB}^\dagger)(O_A \otimes O_B)\big]\big| \le |T_{AB}-T_{AB}^\dagger|_p|O_A|_q|O_B|_q, \quad \frac{1}{p}+\frac{1}{q}=1 \ .
\end{equation}
Similar commutator inequalities also hold true with our GSDK.
Especially for $p=2$, which is particularly significant. A non-vanishing  Schatten 2-norm ($|T_{AB}-T^{\dagger}_{AB}|_{2}\neq 0$) signals that subsystem $A$ at time 0 can influence subsystem $B$ at a later time t. For an arbitrary initial state $\rho$, we may compute the Schatten $2$-norm of this quantity, using the definition \eqref{eq:one}. It is convenient to begin with the identity:
\begin{equation}\label{eq:conv_identity}
    |T_{AB}-T^{\dagger}_{AB}|^2_{2}=2 \operatorname{Tr}\left(T_{AB}T^{\dagger}_{AB}\right)-2\Re \operatorname{Tr}\left(T^{2}_{AB}\right)  \ .
\end{equation}
To illustrate how each term can be evaluated in general, let us consider the first contribution, $\Tr\left(T_{AB}T^{\dagger}_{AB}\right)$ and, for simplicity, take the local subsystems $A$ and $B$ with a finite dimensional Hilbert space. Starting from the definition of the bipartite kernel $T_{AB}(t)$, we obtain
\begin{align}
\operatorname{Tr}\!\big(T_{AB}T_{AB}^\dagger\big)
&=\sum_{i,j,k,l}\sum_{m,n,p,q}
C_{ij;kl}(t) C^{*}_{mn;pq}(t)
\operatorname{Tr}\!\Big[(E^{A}_{ji} \otimes E^{B}_{lk})(E^{A}_{mn} \otimes E^{B}_{pq})\Big] \nonumber \\
&=\sum_{i,j,k,l}\big|C_{ij;kl}(t)\big|^2.
\label{eq:eval_TTdag}
\end{align}
An analogous computation yields the second contribution in \eqref{eq:conv_identity}. Combining the two results, the Schatten 2-norm $|T_{AB}(t)-T_{AB}(t)^\dagger|_2^2$ takes the compact form
\begin{equation}
    |T_{AB}(t)-T_{AB}(t)^\dagger|_2^2
    =2\sum_{i,j,k,l}\Big(|C_{ij;kl}(t)|^2-\Re\big[C_{ij;kl}(t) C_{ji;lk}(t)\big]\Big). 
\label{eq:res_norm_sq}
\end{equation}
These simple equations show that the Schatten $2$-norm is specially convenient for practical calculations of the bounds on temporal correlations. Of course, it is to be expected that all norms will provide essentially the same physical information due to the general inequalities shown above.

Everything here is expressed entirely in the subsystem basis via $C_{ij;kl}(t)$, defined in \eqref{eq:one}. So, the only $t$–dependence enters through the Heisenberg-evolved insertion on $B$.
Consequently, both terms in \eqref{eq:res_norm_sq} inherit this $t$–dependence in general. There are, however, two important situations in which the time dependence drops out:

\textbf{(1) Whole-system specialisation.} If one picks $A=B=S$ (i.e., use the basis on the full system rather than a subsystem, so the operators $O_{S_i}$ are allowed to be any bounded observable), the construction becomes time-independent. Starting with the definition of the $2$-point kernel: 
\begin{equation}
    T^{(2)}(t)=\sum_{i,j,k,l} C_{ij;kl}(t)\; E_{ji}\otimes E_{lk} \ ,
\end{equation}
where $C_{ij;kl}(t)=\operatorname{Tr}\!\big(\rho\;E_{ij}U_t^\dagger E_{kl}U_t\big)$ with a general density matrix $\rho$ (Hermitian, $\rho>0$, $\Tr \rho=1$) and $U=e^{-iHt}$. In this case, \eqref{eq:res_norm_sq} collapses to a constant.

Indeed, one can show $\operatorname{Tr}\left(\left(T^{(2)}(t)\right)^2\right)=\big(\operatorname{Tr}\rho\big)^2$ and $\operatorname{Tr}\left(T^{(2)}(t)T^{(2)\dagger}(t)\right)=L\operatorname{Tr}\rho^2$, so \eqref{eq:res_norm_sq} is time-independent. To demonstrate how this works, we proceed with a calculation of the Schatten $p$-norms in the case where the GSDK is defined with respect to a thermal state $\rho_\beta$. For the two-point kernel $T_{AB}(t)$, an analytical form is known from the work in \cite{Milekhin:2025ycm}, but we may also use the formalism developed up to this point, in which case, to compute the coefficients $C^{(\beta)}_{ij;kl}(t)$, we apply the matrix-unit conjugation rule $U_t^\dagger E_{kl}U_t=\sum_{a,b}U^*_{ka}U_{lb}E_{ab}$ and the fact that $\rho_\beta E_{ij}=(\rho_\beta)_iE_{ij}$, as $\rho_\beta$ is diagonal in the energy basis. These imply, together with the identity $\operatorname{Tr}(E_{ij}E_{ab})=\delta_{ja}\delta_{ib}$, that
\begin{align}\label{eq:c_beta_eval}
C^{(\beta)}_{ij;kl}(t)
&= \operatorname{Tr}\!\left[((\rho_\beta)_iE_{ij})\sum_{a,b}U^*_{ka}U_{lb}((\rho_\beta)_aE_{ab})\right] \\
&= (\rho_\beta)_i(\rho_\beta)_jU^*_{kj}U_{li}. \nonumber
\end{align}

From this point, simple analytical manipulations lead to the identity
\begin{equation}\label{eq:t2_norm_y2_norm}
 |T_{AB}(t)|_p=|\rho_\beta|_p^{2} = \left[\frac{Z(p\beta)}{Z(\beta)^p}\right]^\frac{2}{p}.
\end{equation}

In addition, an analogous calculation for $T^{(4)}(t)$, detailed in the Appendix \ref{appendix:B}, yields 
\begin{equation}\label{eq:y_norm_eval}
|T^{(4)}(t)|_p = \left[\frac{Z\!\left(\tfrac{p}{4}\beta\right)}{Z(\beta)^{p/4}}\right]^\frac{4}{p},
\end{equation}
and similar results are also valid for other $T^{(n)}(t_1,...,t_n)$. 

More interesting than the expressions themselves, which are further manifestations of the connection between the GSDK and SFF established in Eqs. (\ref{eq:trace_relations}), is their \textit{time-independence}, despite this not being the case for the operators $T^{(n)}(t_1,...,t_n)$ in general. After all, this constant behaviour is valid for all possible time evolutions, including drastically different dynamics such as integrability and chaos, so how to understand it?

The answer comes directly from the specification of these kernels $T^{(n)}$ to the full quantum system: the time evolution $O(t)$ of some observable $O$ is, trivially, also an operator $O'$ with which correlation functions may be calculated. Thus, if the correlators of all observables of the system may be derived using $T^{(n)}$, the bound provided by the norms at a given time slice must remain exactly the same at all others. From the point of view of using the GSDKs to identify chaos, the constant $p$-norm of the kernel $\tilde{T}^{(4)}$ which generates all thermal-regulated OTOCs can also be understood as a consequence of the fact that chaotic behaviour in a quantum system is be characterized by the scrambling of ```simple" operators at $t=0$. It is now understood that, whatever the proper definition of such observables is, it includes all \textit{local} operators \cite{Maldacena:2015waa, DAlessio:2015qtq} and it is the OTOCs built from those that show interesting time evolutions. With this in mind, it is no surprise that the norm of a GSDK which takes all possible operators as input is ```trivial", as it does not discriminate between simple and complex operators. 

Overall, this general time-independence points to the study of kernels restricted to local subsystems as a potentially fruitful line of further investigation.

\textbf{(2) Stationary states.} If $[\rho,U_t]=0$ for all $t$ (e.g., $\rho=f(H)$, including thermal states) in the global basis, then the unitary conjugations on the $B$ leg can be moved past $\rho$ in each trace and cancel pairwise inside the basis-completeness contractions; the resulting sums in \eqref{eq:res_norm_sq} become independent of $t$. This is the same mechanism that makes the $\beta$-regularised 2- and 4-leg constructions reduce to spectral-form-factor expressions in the energy basis; the cancellations come from cyclicity of trace and the matrix-unit identities.

Except these special cases, nothing forces the phases and rotations from $U_t^{\dagger}\widehat E^B_{kl}U_t$ to cancel inside the double sums of \eqref{eq:res_norm_sq}, so $|T_{AB}(t)-T_{AB}(t)^\dagger|_2$ is generically time-dependent for a general density matrix in a genuine subsystem basis. For the remainder of this section, however, the focus lies on the Schatten norms $|T^{(n)}(t_1,...,t_n)|_p$, which we explore with a couple of examples.


\subsection{Kernel norms and the Eigenstate Thermalization Hypothesis}\label{sec:ETH bound}

Given the overall discussion on bounds of correlation functions and this work's interest in exploring connections to quantum chaos, it becomes essential to establish the relation between the spacetime density kernel and the seminal ``bound on chaos" first derived in \cite{Maldacena:2015waa}, where the maximal exponent for the intermediary time growth rate of OTOCs was found.

As explained in Section \ref{sec:intro}, the 4-leg kernel $T^{(4)}(t)$ satisfies, for any operator $A$ acting on a subsystem
\begin{equation} \label{eq:5}
\operatorname{Tr}\!\left[T^{(4)}(t)(A\otimes A\otimes A\otimes A)\right] = \operatorname{Tr}\!\left[yA\,yA(t)\,yA\,yA(t)\right] =: \mathcal{T}_A(t),
\end{equation}
and thus reproduces the standard equal-spacing-regulated OTOC (see Eq. (\ref{eq:otoc3}) with $y^4=\rho$ and all observables chosen as $A$), with the notation $\mathcal{T}_A(t)$ introduced for convenience. 

A brilliant study of the OTOC $\mathcal{T}_A(t)$ was made by \cite{Murthy:2019fgs}, where the bound on the growth exponent was deduced using as starting point the Eigenstate Thermalization Hypothesis (ETH), nowadays understood to correctly describe the behaviour of ``simple observables" (which, in the context of many-body physics include local operators) in chaotic \cite{DAlessio:2015qtq}, along with other smaller inputs from studies in holography. As the derivation is valid for any choice of local $A$, it once again suggests that it should be possible to characterize the chaotic aspects of a quantum system directly from a GSDK such as $T^{(4)}(t)$, with bounds like Eq. (\ref{eq:gsdk_bound1}) extending the result to all OTOCs and the work of \cite{Murthy:2019fgs} indicating that the way forward is through an application of the ETH.

With this in mind, we now sketch the main steps of the calculation in \cite{Murthy:2019fgs} before translating them to $T^{(4)}(t)$ (more specifically, its Schatten $p$-norm). As we will see, $\mathcal{T}_A(t)$ is naturally split into disconnected (products of regulated two-points) and connected parts as $\mathcal{T}_A(t)=\mathcal{T}_{A,\mathrm{disc}}(t)+\mathcal{T}_{A,c}(t)$, with the latter component being the one subject to the bound on chaos. To begin with, expand $\mathcal{T}_A(t)$ by inserting four resolutions of identity in the energy basis. Using $[y,U_t]=0$ and Heisenberg evolution $A(t)_{ij}=e^{i(E_i-E_j)t}A_{ij}$), one finds
\begin{equation} \label{eq:7}
\begin{aligned}
\mathcal{T}_A(t) &=\sum_{i,j,k,l} (y_i y_j) (y_j y_k) (y_k y_l) (y_l y_i) e^{i[(E_j-E_k)+(E_l-E_i)]t} A_{ij}A_{jk}A_{kl}A_{li}\\
&=\sum_{i,j,k,l} y_i^2 y_j^2 y_k^2 y_l^2 e^{i(\omega_2-\omega_1)t} A_{ij}A_{jk}A_{kl}A_{li},
\end{aligned}
\end{equation}
with $\omega_1:=E_i-E_j$, $\omega_2:=E_k-E_l$ (and note that $E_j-E_k= -\omega_2+\omega_1$, $E_l-E_i= -\omega_1$).

Now, the ETH is applied for the matrix elements of $A_{ij}$ and we assume that the thermal mean $A$ has been subtracted, which leads to \cite{Murthy:2019fgs, DAlessio:2015qtq}
\begin{equation} \label{eq:6}
A_{ij}=e^{-\frac{S(E)}{2}}f(E,\omega)R_{ij},\qquad E=\frac{E_i+E_j}{2},\ \omega=E_i-E_j,
\end{equation}
where $f(E,\omega)$ is smooth and even in ($\omega$), and the random matrix $R_{ij}$ has zero mean and unit variance on local energy shells. 

Replacing this expression in place of each $A_{ij}$ in Eq. (\ref{eq:7}) and taking the statistical average of the product of four $R$-matrices, the work of Foini, Kurchan and Pappalardi \cite{Foini:2018sdb, Pappalardi:2022aaz} proved that this results in
\begin{equation} \label{eq:8}
R_{ij}R_{jk}R_{kl}R_{li}=\delta_{ik}+\delta_{jl}+e^{-S(E)}g(E,\omega_1,\omega_2,\omega_3).
\end{equation}

Thus, the Kronecker deltas generate the disconnected pieces (products of two regulated two-point functions), while the last term gives the connected OTOC part $\mathcal{T}_{A,c}(t)$. At this point, the analysis consists of combining Eqs. (\ref{eq:7}) and (\ref{eq:8}), replacing the energy sums with integrals through $\sum_i\to\int dE_i\,e^{S(E_i)}$, taking a Fourier transform to frequency space and properly manipulating the resulting expression along the lines detailed in \cite{Murthy:2019fgs}. It is interesting to point out that up to this point, it is possible to derive the inequality, for large values of $|\omega|$,
\begin{equation} \label{eq:18}
 |\widetilde{\mathcal{T}}_{A,c}(\omega)| \lesssim e^{-\frac{3\beta}{4}|\omega|}.
\end{equation}
Through the inverse Fourier transform (and the general results in Fourier analysis \cite{steinfourier}), this implies that $\mathcal{T}_{A,c}(t)$ is $C^\infty$ and is analytic within a strip of width $\frac{3\beta}{4}$ and, when the bound is saturated, it decays algebraically. From this a Lyapunov exponent for the OTOC may be derived with the help of additional (necessary) assumptions such as time-reversal symmetry as well as analytical structures drawn from holography, which lead to an upper bound on the exponent of $\frac{4\pi}{3\beta}$ \cite{Murthy:2019fgs}. In Appendix~\ref{appendix:C} we recast this ETH-based analysis directly
at the level of the four-leg kernel. By working with the thermal inner product
and the projector onto the subspace of local operators, we define a local norm
for the connected kernel $T^{(4)}_c(t)$ and show that the
bound in \cite{Murthy:2019fgs} is equivalent to the operator inequality \eqref{eq:42}. This makes the chaos bound a operator constraint on the local sector of the generalized spacetime density kernel and, conversely, it might be possible to use the GSDK to characterize the bound on chaos.

\section{Discussions}\label{sec:discussion}

In this work, we have developed a unified framework for the time-like entanglement proposal of \cite{Milekhin:2025ycm}, especially using a local/global orthonormal basis and extending to a natural generalization to what we have referred to as the generalized spacetime density kernel (GSDK).
For 2N-point functions this leads naturally to a 
2N-leg GSDK, which may be regarded as a dynamical extension of a density matrix for temporally extended subsystems with support on more than one Cauchy slices.

A central outcome of this construction is a precise relation between the GSDK and spectral diagnostics of chaos. We have explicitly demonstrated how the higher point function ({\it e.g.}~four-point function and, in general, a $(2N)$-point function) produces higher moments of the corresponding SFF, evaluated at an effective $N$-enhanced temperature. This provides a bridge between early-time scrambling, as probed by OTOCs, and late-time random-matrix universality encoded in the slope–dip–ramp–plateau structure of the spectral form factor. The GSDK thus furnishes a single object that simultaneously generates higher-point temporal correlators and while capturing the spectral information. We have also explicitly demonstrated how several generic bounds, including those from the eigenstate thermalization hypothesis, act on the GSDK operator. These results explicitly demonstrate the usefulness of the local orthonormal basis.

Given the generic and universal nature of the GSDK  operator, we have barely scratched the surface, and there are several natural and interesting aspects to investigate further. For example, it remains to be seen how efficiently the GSDK operator can capture and demarcate between the dynamical aspects of a quantum system. Especially since the higher point generalizations are directly related to the higher moments of the SFF. Work in this direction is in progress.

Similarly to the proposal in \cite{Milekhin:2025ycm}, the GSDK operator is defined for a finite dimensional Hilbert space, even though we have ignored the subteties and discusses examples of CFT in the context of SFF and its behaviour. It would be very interesting to extend the definition for QFTs, especially for CFTs, and properly understand its relation to modular theory. Given the basic structures, it is expected that the GSDK operators are analogous to the higher moments of the density matrix and, therefore, presumably extend these notions beyond Renyi entropies, suggesting that one possible application may be the definition of ``multipartite timelike entanglement". 

Given the multitude of definitions of timelike entanglement, it will certainly be interesting to connect these notions systematically.\footnote{\cite{Milekhin:2025ycm} has already discussed some of these connections.} Of particular interest are perhaps the Holographic CFTs, for which {\it a} notion of timelike entanglement is simply defined based on a straightforward analytic continuation of the usual entanglement entropy of a given sub-system\cite{Nakata:2020luh, Mollabashi:2020yie, Mollabashi:2021xsd, Caputa:2024gve}.\footnote{Relatedly, it is important to also understand whether such timelike entanglement measures are true entanglement monotones.} We hope to address some of these issues in near future.

\section{Acknowledgements}

We thank Alexey Milekhin, Krishanu Roychowdhury, Ashish Shukla, Tadashi Takayanagi for discussions and conversations on related topics. A special thanks goes to Ashish Shukla for emphasizing the importance of timelike entanglement and collaboration at the initial stages of this work. RND and MHMC also thank Tadashi Takayanagi and Jonathan Harper for insightful discussions on the matter. AK acknowledges the support of the Humboldt Research Fellowship for Experienced Researchers by the Alexander von Humboldt
Foundation and for the hospitality of Theoretical Physics III, Department of Physics and Astronomy, Julius-Maximilians-Universit\"{a}t W\"{u}rzburg and the support from the ICTP through the Associates Programme (2024-2030) during the course of this work. RND's work leading to this publication was supported by the PRIME programme of the German Academic Exchange Service (DAAD) with funds from the German Federal Ministry of Research, Technology and Space (BMFTR).  RND is also supported by Germany's Excellence Strategy through the W\"urzburg‐Dresden Cluster of Excellence on Complexity and Topology in Quantum Matter ‐ ct.qmat (EXC 2147, project‐id 390858490).

\appendix

\section{Proof of the Haar averaging of the operators}\label{appendix:A}

Here we review the derivation of Haar averaging of $V\otimes V^\dagger$ for unitary operator $V$ that results in the SWAP operator. Let $\mathcal H\cong\mathbb C^L$ with orthonormal basis $\{|i\rangle\}_{i=1}^L$. For Haar-random $V\in U(L)$ we want to show
\begin{equation}
    \ \mathbb{E}_{V\sim \mathrm{Haar}}\,[\,V\otimes V^\dagger\,]=\frac{1}{L}\,\mathrm{SWAP}\, \label{eq:goal}
\end{equation}
where the SWAP operator on $\mathcal H\otimes\mathcal H$ is $\mathrm{SWAP}(|i\rangle\otimes|j\rangle)=|j\rangle\otimes|i\rangle$, equivalently $\mathrm{SWAP}=\sum_{i,j=1}^L E_{ij}\otimes E_{ji}$,
and $E_{ij}:=|i\rangle\!\langle j|$ are the matrix units. We will first prove the Haar first moment (twirl) $\displaystyle \int UOU^\dagger\,d\mu(U)=\frac{\mathrm{Tr}(O)}{L}\,I$. Then convert it into the index identity $\displaystyle \int U_{jk}U^*_{m\ell}\,d\mu(U)=\frac{1}{L}\delta_{jm}\delta_{k\ell}$ and finally, use it to compute $\mathbb E[V\otimes V^\dagger]=\frac{1}{L}\mathrm{SWAP}$.
Throughout, $d\mu(U)$ denotes the (normalized) Haar probability measure on $U(L)$. Every matrix $A$ expands as $A=\sum_{i,j}A_{ij}E_{ij}$. For any basis vectors $|a\rangle\otimes|b\rangle$,
\begin{align}
    \Big(\sum_{i,j}E_{ij}\otimes E_{ji}\Big)\,(|a\rangle\otimes|b\rangle)
    =|b\rangle\otimes|a\rangle.
\end{align}
The Haar twirl, a linear map on operators is defined as
\begin{equation}
    \Phi(O):=\int_{U(L)} UOU^\dagger\,d\mu(U). \label{eq:twirl_def}
\end{equation}
We choose a $V\in U(L)$. Using left-invariance of Haar measure,
\begin{align}
    V\,\Phi(O)\,V^\dagger
    &=\int VU\,O\,U^\dagger V^\dagger\,d\mu(U)
    =\int W\,O\,W^\dagger\,d\mu(W)\quad(W=VU) \nonumber \\
    &=\Phi(O).
\end{align}
Hence $V\Phi(O)=\Phi(O)V$ for all unitaries $V$. Let $Y$ satisfy $VY=YV$ for all $V\in U(L)$. In the basis $\{|i\rangle\}$:
For any diagonal phase $D_\theta=\mathrm{diag}(e^{i\theta_1},\dots,e^{i\theta_L})$, from $D_\theta Y=Y D_\theta$ we get
$(e^{i\theta_i}-e^{i\theta_j})Y_{ij}=0$ for all $\theta$, hence $Y_{ij}=0$ if $i\neq j$.
For any permutation matrix $P$, from $PY=YP$ we get $Y_{11}=Y_{22}=\cdots=Y_{LL}$.
So $Y=c\,I$ for some scalar $c$.
$\mathrm{Tr}\,\Phi(O)=\int \mathrm{Tr}(UOU^\dagger)\,d\mu(U)=\mathrm{Tr}(O)$ by cyclicity. If $\Phi(O)=cI$, then $\mathrm{Tr}\,\Phi(O)=c\,\mathrm{Tr}(I)=cL$. Therefore
\begin{equation}
    \ \Phi(O)=\frac{\mathrm{Tr}(O)}{L}\,I. \label{eq:twirl_result}
\end{equation}
This is the Haar first-moment identity. Next, expand $UE_{k\ell}U^\dagger$ entirely in matrix units,
\begin{equation}
    U=\sum_{p,q}U_{pq}\,E_{pq},\qquad
    U^\dagger=\sum_{r,s}U^*_{sr}\,E_{rs}.
\end{equation}
We use the identities $U\,E_{k\ell}
    =\sum_{p}U_{pk}\,E_{p\ell}$ and $UE_{k\ell}U^\dagger=\sum_{j,m}U_{jk}\,U^*_{m\ell}\,E_{jm}$ and average both sides with Haar measure and use the twirl
\begin{equation}
    \int UE_{k\ell}U^\dagger\,d\mu(U)
    =\frac{\mathrm{Tr}(E_{k\ell})}{L}\,I
    =\frac{\delta_{k\ell}}{L}\sum_{j}E_{jj}.
\end{equation}
Performing the averaging explicitly gives
\begin{equation}
    \int UE_{k\ell}U^\dagger\,d\mu(U)
    =\sum_{j,m}\Big(\int U_{jk}U^*_{m\ell}\,d\mu(U)\Big)\,E_{jm}.
\end{equation}
Because the $E_{jm}$ form a basis, the coefficients of each $E_{jm}$ must match. Therefore
\begin{equation}
     \int_{U(L)} U_{jk}\,U^*_{m\ell}\,d\mu(U)
    =\frac{1}{L}\,\delta_{jm}\,\delta_{k\ell}.\label{eq:index}
\end{equation}
Finally we compute $\mathbb E[V\otimes V^\dagger]$. Expand $V$ and $V^\dagger$ in matrix units $V=\sum_{i,k}V_{ik}E_{ik},
    V^\dagger=\sum_{j,\ell}V^*_{j\ell}E_{\ell j}$.
Then $V\otimes V^\dagger
    =\sum_{i,k}\sum_{j,\ell}V_{ik}\,V^*_{j\ell}\,(E_{ik}\otimes E_{\ell j})$.
Averaging with Haar and using the index identity of \eqref{eq:index}finally gives
\begin{align}
    \int V\otimes V^\dagger\,d\mu(V)
    &=\sum_{i,k}\sum_{j,\ell}\Big(\int V_{ik}V^*_{j\ell}\,d\mu(V)\Big)\,(E_{ik}\otimes E_{\ell j}) \nonumber \\
    &=\sum_{i,k}\sum_{j,\ell}\frac{1}{L}\,\delta_{ij}\delta_{k\ell}\,(E_{ik}\otimes E_{\ell j}) \nonumber \\
    &=\frac{1}{L}\sum_{i,k} E_{ik}\otimes E_{ki}
    =\frac{1}{L}\,\mathrm{SWAP}.
\end{align}
This is the identity we use
\begin{equation}
\mathbb{E}_{V\sim\mathrm{Haar}}\,[\,V\otimes V^\dagger\,]=\frac{1}{L}\,\mathrm{SWAP}. \label{eq:final_result}
\end{equation}

\section{Constant Schatten \texorpdfstring{$p$}{p}-norm of 
         \texorpdfstring{$T^{(4)}(t)$}{T4(t)}}
\label{appendix:B}
We work on a finite-dimensional Hilbert space $\mathcal{S}$ of size $L$, with the matrix unit basis $E_{ij}=|i\rangle\langle j|$. We will repeatedly use the following identities
\begin{equation}\label{eq:matrix_unit_rules}
\operatorname{Tr}(E_{ji}X)=X_{ij},\qquad E_{ab}E_{cd}=\delta_{bc}\,E_{ad}.
\end{equation}
For arbitrary probes $O_1,\dots,O_4$, we define
\begin{equation}\label{eq:ft_def_new}
\mathcal{F}_t[O_1,O_2,O_3,O_4] =\operatorname{Tr}\!\big[y\,O_1\,y\,U_t^\dagger O_2U_t\,y\,O_3\,y\,U_t^\dagger O_4U_t\big],
\end{equation}
and construct a single 4-leg operator $T^{(4)}(t)$ such that
\begin{equation}\label{eq:t4_def_new}
\mathcal{F}_t[O_1,O_2,O_3,O_4]=\operatorname{Tr}\!\left[T^{(4)}(t)(O_1\!\otimes\!O_2\!\otimes\!O_3\!\otimes\!O_4)\right].
\end{equation}
Expanding each probe in matrix units and grouping coefficients yields the numbers
\begin{equation}\label{eq:c_def}
C_{ij;kl;mn;pq}(t):=\operatorname{Tr}\!\left[y\,E_{ij}\,y\,(U_t^\dagger E_{kl}U_t)\,y\,E_{mn}\,y\,(U_t^\dagger E_{pq}U_t)\right]
\end{equation}
so that the operator can be written as
\begin{equation}\label{eq:t4_expansion}
T^{(4)}(t)=\sum_{i,j,k,l,m,n,p,q} C_{ij;kl;mn;pq}(t)\; E^{(1)}_{ji}\!\otimes E^{(2)}_{lk}\!\otimes E^{(3)}_{nm}\!\otimes E^{(4)}_{qp}.
\end{equation}
We will also use two standard expansions:
\begin{equation}\label{eq:unitary_conj}
U E_{k\ell}U^\dagger=\sum_{j,m}U_{jk}U^*_{m\ell}\,E_{jm},
\end{equation}
and in the energy basis,
\begin{equation}\label{eq:y_action}
yE_{ab}y=y_ay_b\,E_{ab}.
\end{equation}
We now evaluate the coefficients $C_{ij;kl;mn;pq}(t)$ explicitly. Inserting the matrix-unit expansions from \eqref{eq:unitary_conj} and the diagonal action of $y$ from \eqref{eq:y_action} into \eqref{eq:c_def}, we have
\begin{align}
C_{ij;kl;mn;pq} &= \operatorname{Tr}\!\left[(y_i y_j)E_{ij} \left(\sum_{a,b}U^*_{ka}U_{lb}E_{ab}\right) (y_m y_n)E_{mn} \left(\sum_{c,d}U^*_{pc}U_{qd}E_{cd}\right)\right] \\
&= \sum_{a,b,c,d}y_i y_j y_m y_n\;U^*_{ka}U_{lb}U^*_{pc}U_{qd}\; \operatorname{Tr}\!\big(E_{ij}E_{ab}E_{mn}E_{cd}\big).
\end{align}
The trace of the product of matrix units is evaluated using \eqref{eq:matrix_unit_rules}:
\begin{equation}
\operatorname{Tr}(E_{ij}E_{ab}E_{mn}E_{cd})=\delta_{ja}\,\delta_{bm}\,\delta_{nc}\,\delta_{id}.
\end{equation}
Plugging this back into the sum and performing the summation over $a, b, c, d$ collapses the expression to:
\begin{equation}\label{eq:c_result}
C_{ij;kl;mn;pq} = y_i y_j y_m y_n\;U^*_{k j}\,U_{l m}\,U^*_{p n}\,U_{q i}.
\end{equation}
To write this more compactly, we introduce two $L\times L$ matrices
\begin{equation}\label{eq:m1_def}
M_1:=yU_t^\dagger \quad \text{with components} \quad (M_1)_{jk}=y_jU^*_{kj},
\end{equation}
and
\begin{equation}\label{eq:m2_def}
M_2:=U_t y \quad \text{with components} \quad (M_2)_{lm}=U_{lm}\,y_m.
\end{equation}
Then the coefficients in \eqref{eq:c_result} can be rewritten as
\begin{equation}\label{eq:c_compact}
C_{ij;kl;mn;pq}=(M_1)_{jk}\,(M_2)_{lm}\,(M_1)_{np}\,(M_2)_{qi}.
\end{equation}
The operator $T^{(4)}(t)$ is written in the basis $\{E^{(1)}_{ji}\!\otimes E^{(2)}_{lk}\!\otimes E^{(3)}_{nm}\!\otimes E^{(4)}_{qp}\}$. If we regard $T^{(4)}(t)$ as an $L^4\times L^4$ matrix with row multi-index $(j,l,n,q)$ and column multi-index $(i,k,m,p)$, then by definition the matrix element is
\begin{equation}\label{eq:t4_matrix_elem}
\big[T^{(4)}(t)\big]_{(j,l,n,q),(i,k,m,p)} = C_{ij;kl;mn;pq}.
\end{equation}
Using the compact form from \eqref{eq:c_compact}, this becomes
\begin{equation}\label{eq:t4_matrix_elem_compact}
\big[T^{(4)}(t)\big]_{(j,l,n,q),(i,k,m,p)} = \underbrace{(M_1)_{j,k}}_{\text{pairs }(j,k)}\, \underbrace{(M_2)_{l,m}}_{\text{pairs }(l,m)}\, \underbrace{(M_1)_{n,p}}_{\text{pairs }(n,p)}\, \underbrace{(M_2)_{q,i}}_{\text{pairs }(q,i)}.
\end{equation}
We compare this with the 4-fold Kronecker product
\begin{equation}\label{eq:k_def}
K(t):=M_1\otimes M_2\otimes M_1\otimes M_2,
\end{equation}
whose matrix elements are
\begin{equation}\label{eq:k_matrix_elem}
\big[K(t)\big]_{(j,l,n,q),(i,k,m,p)} =(M_1)_{j,i}\,(M_2)_{l,k}\,(M_1)_{n,m}\,(M_2)_{q,p}.
\end{equation}
Equations \eqref{eq:t4_matrix_elem_compact} and \eqref{eq:k_matrix_elem} differ only by a reordering of the column multi-index $(i,k,m,p)\mapsto (k,m,p,i)$. Let $\Pi_R$ be the $L^4\times L^4$ permutation matrix that implements this reordering on column indices. This gives the exact identity
\begin{equation}\label{eq:t4_is_k_pi}
\qquad T^{(4)}(t)\;=\;K(t)\,\Pi_R.\qquad
\end{equation}
Because $\Pi_R$ is a permutation matrix, it is unitary, and thus left or right multiplication by it does not change singular values. Therefore all Schatten norms agree:
\begin{equation}\label{eq:norm_t_equals_norm_k}
 |T^{(4)}(t)|_p=|K(t)|_p\ \ \text{for all }p\in[1,\infty].\
\end{equation}
We can now evaluate the norm of $K(t)$ using the multiplicativity of singular values under Kronecker products and the unitary invariance of Schatten norms
\begin{equation}\label{eq:norm_k_eval}
|K(t)|_p=|M_1|_p\,|M_2|_p\,|M_1|_p\,|M_2|_p =|M_1|_p^2\,|M_2|_p^2.
\end{equation}
Since $M_1=yU_t^\dagger$ and $M_2=U_t y$, we have $|M_1|_p=|M_2|_p=|y|_p$. Hence, from \eqref{eq:norm_t_equals_norm_k},
\begin{equation}\label{eq:t4_norm_result}
 |T^{(4)}(t)|_p=|y|_p^{\,4}.\ 
\end{equation}
We can write $y=\rho_\beta^{1/4}$ with $\rho_\beta=e^{-\beta H}/Z(\beta)$. The p-norm of y is then
\begin{equation}\label{eq:y_norm_eval_app}
|y|_p^p=\operatorname{Tr}(y^p)=\operatorname{Tr}\!\big(\rho_\beta^{p/4}\big) =\frac{Z\!\left(\tfrac{p}{4}\beta\right)}{Z(\beta)^{p/4}},
\end{equation}
where $Z(\beta)=\operatorname{Tr}e^{-\beta H}$. This gives the final closed form
\begin{equation}\label{eq:t4_norm_final_app}
 |T^{(4)}(t)|_p =\Big(\frac{Z(\tfrac{p}{4}\beta)}{Z(\beta)^{p/4}}\Big)^{\!4/p}.
\end{equation}
This result shows that $|T^{(4)}(t)|_p$ is independent of time $t$.

\section{A lower bound on the Schatten 
         \texorpdfstring{($p$)}{(p)}-norm of 
         \texorpdfstring{$T^{(4)}(t)$}{T4(t)}}
\label{appendix:C}

A lower-bound for the Schatten-$p$ norm of the 4-leg kernel, $|T^{(4)}(t)|_p$, can be derived directly from H\"older-type inequalities and an explicit 4-cycle SWAP operator, which we will name a ```tester", as its properties make it specially suitable for certain analyses. The four-point OTOC functional is given by
\begin{equation}\label{eq:otoc_functional}
\mathcal{F}_t[O_1,O_2,O_3,O_4]=\operatorname{Tr}\!\left[yO_1yU_t^\dagger O_2U_tyO_3yU_t^\dagger O_4U_t\right]
\end{equation}

A special ``tester'' operator, the 4-cycle $\Omega_{(1234)}$, the 4-leg analogue of SWAP, is defined as
\begin{equation}\label{eq:4_cycle_def}
\Omega_{(1234)}=\sum_{a,b,c,d} E^{(1)}_{ab}\!\otimes E^{(2)}_{bc}\!\otimes E^{(3)}_{cd}\!\otimes E^{(4)}_{da}.
\end{equation}
 Contracting $T^{(4)}(t)$ with $\Omega_{(1234)}$ gives
\begin{equation}\label{eq:f4t_def}
\overline{F_4(t)}\equiv \frac{1}{L}\operatorname{Tr}\!\big[T^{(4)}(t)\Omega_{(1234)}\big]
= \frac{1}{L}\big|\operatorname{Tr}(yU_t)\big|^{4}
= \frac{1}{L}\frac{|Z(\beta/4,t)|^{4}}{Z(\beta)}.
\end{equation}
Here $L=\dim \mathcal{S}$ and $y^4=\rho_\beta=e^{-\beta H}/Z(\beta)$. A general variational lower bound is obtained
\begin{equation}\label{eq:rearranged_bound}
\frac{|\mathcal{F}_t[O_1,O_2,O_3,O_4]|}{\prod_r|O_r|_q}\ \le\ |T^{(4)}(t)|_p.
\end{equation}
Taking the supremum over all operators $O_r$ with $|O_r|_q=1$ gives the best lower bound from factorized probes
\begin{equation}\label{eq:variational_lower_bound}
|T^{(4)}(t)|_p\ \ge\ \sup_{|O_r|_q=1}\ |\mathcal{F}_t[O_1,O_2,O_3,O_4]|
\end{equation}
where $1/p+1/q=1$. The true dual characterization is $|T|_p=\sup_{|X|_q=1}|\operatorname{Tr}(TX)|$; restricting $X$ to factorized tensors only yields a lower bound.

A concrete computable lower bound uses the 4-cycle tester. Applying H\"older with $M=T^{(4)}(t)$ and $N=\Omega_{(1234)}$
\begin{equation}\label{eq:holder_with_omega}
|\operatorname{Tr}[T^{(4)}(t)\Omega]|\ \le\ |T^{(4)}(t)|_p|\Omega|_q,\qquad \frac{1}{p}+\frac{1}{q}=1.
\end{equation}
From \eqref{eq:f4t_def}, $|\operatorname{Tr}[T^{(4)}(t)\Omega]|=L|\overline{F_4(t)}|$. Thus,
\begin{equation}\label{eq:lower_bound_intermediate}
|T^{(4)}(t)|_p\ \ge\ \frac{L|\overline{F_4(t)}|}{|\Omega|_q}.
\end{equation}
The 4-cycle $\Omega$ is a permutation unitary on $\mathcal{S}^{\otimes 4}$, so its singular values are all 1, and
\begin{equation}\label{eq:omega_norm}
|\Omega|_q=\big(\operatorname{Tr} I_{\mathcal{S}^{\otimes 4}}\big)^{1/q}=(L^4)^{1/q}=L^{4/q}.
\end{equation}
Using the explicit $\overline{F_4(t)}$ from \eqref{eq:f4t_def},
\begin{equation}\label{eq:f4t_explicit}
L|\overline{F_4(t)}|=\frac{|Z(\beta/4,t)|^4}{Z(\beta)}.
\end{equation}
This yields the explicit lower bound
\begin{equation}\label{eq:final_lower_bound}
|T^{(4)}(t)|_p\ \ge\ \frac{|Z(\beta/4,t)|^{4}}{Z(\beta)}L^{-4/q}
=\frac{|Z(\beta/4,t)|^{4}}{Z(\beta)}L^{-(4-4/p)}
\end{equation}
with $1/p+1/q=1$. For special cases: for $p=1$ ($q=\infty$), $|\Omega|_\infty=1$, implying $|T^{(4)}(t)|_{1}\ge \frac{|Z(\beta/4,t)|^{4}}{Z(\beta)}$. For $p=2$ ($q=2$), $|\Omega|_2=L^2$, implying $|T^{(4)}(t)|_{2}\ge \frac{|Z(\beta/4,t)|^{4}}{Z(\beta)}L^{-2}$. For $p=\infty$ ($q=1$), $|\Omega|_1=L^4$, implying $|T^{(4)}(t)|_{\infty}\ge \frac{|Z(\beta/4,t)|^{4}}{Z(\beta)}L^{-4}$.
It can be shown that (App.~\ref{appendix:B})
\begin{equation}\label{eq:t4_norm_final}
 |T^{(4)}(t)|_p =\Big(\frac{Z(\tfrac{p}{4}\beta)}{Z(\beta)^{p/4}}\Big)^{\!4/p}.
\end{equation}
which is independent of time. So this bound is \textit{trivially true}. $|T^{(4)}(t)|_p$ can be seen as an upper bound of the 4-point SFF with a quarter thermal weight.

\paragraph{The operator growth bound for the connected kernel} In our reproduction of the ETH bound on chaos result from \cite{Murthy:2019fgs}, if one had dropped an extra ($e^{-\beta|\omega-\omega_1|/4}$) coming from ($g$), the best lower bound on the sum would be ($2|\omega|$), leading to a weaker ($e^{-\beta|\omega|/2}$). The connector is essential for the optimal ($3\beta/4$) slope. 

This is precisely the ETH-based derivation of the MSS bound for the OTO connected part in this family. (Eqs. (28)--(30) in \cite{Murthy:2019fgs} show the steps; the Paley--Wiener condition they give guarantees an extra ($e^{-(\pi/2\lambda)|\omega|}$) from ($K(\omega)$).) We now phrase the result at the operator level for the kernel ($T^{(4)}(t)$) itself. Define the thermal inner product on operators ($X,Y$) on ($\mathcal{S}$) by 
\begin{equation} \label{eq:31}
\langle X,Y\rangle_\beta := \operatorname{Tr}\!\left(\rho_\beta^{1/2}X^\dagger \rho_\beta^{1/2}Y\right) = \operatorname{Tr}\!\left(y^2 X^\dagger y^2 Y\right).
\end{equation}
A convenient way to work with it is to isometrically embed the operator space into the ordinary HS space by the similarity transform
\begin{equation} \label{eq:32}
\Phi:\ X\mapsto \tilde{X}:=y\,X\,y, \qquad \langle X,Y\rangle_\beta=\operatorname{Tr}\left(\tilde{X}^\dagger \tilde{Y}\right).
\end{equation}
Let (\{$B_a$\}) be any local operator basis that is orthonormal in the thermal HS sense,
\begin{equation} \label{eq:33}
\langle B_a,B_b\rangle_\beta=\delta_{ab}=\operatorname{Tr}\left(y^2 B_a^\dagger y^2 B_b\right).
\end{equation}
Then the orthogonal projector onto the local subspace is the rank-($N_{\text{loc}}$) superoperator
\begin{equation} \label{eq:34}
P_{\mathrm{loc}}(X)=\sum_{a=1}^{N_{\text{loc}}}\langle B_a,X\rangle_\beta B_a = \sum_a \operatorname{Tr}\!\left(y^2 B_a^\dagger y^2 X\right) B_a.
\end{equation}
Equivalently, in ``Liouville-ket'' notation ($|X\rrangle$) for operators with inner product ($\llangle X|Y\rrangle_\beta=\langle X,Y\rangle_\beta$), one has
\begin{equation} \label{eq:35}
P_{\mathrm{loc}}=\sum_a |B_a\rrangle \llangle B_a| .
\end{equation}
For the four-leg space ($( \text{Op}(\mathcal{S}) )^{\otimes 4}$), we use the legwise projector
\begin{equation} \label{eq:36}
 P_{\mathrm{loc}}^{(4)}:=P_{\mathrm{loc}}^{\otimes 4}.
\end{equation}
If an operator $X$ is denoted by $|X\rangle\!\rangle$. The inner product in this space is defined as
\begin{equation}
\label{eq:inner_product}
\langle\!\langle X|Y\rangle\!\rangle_\beta = \langle X,Y\rangle_\beta = \mathrm{Tr}(y^2 X^\dagger y^2 Y).
\end{equation}
Using a local orthonormal basis $\{B_a\}$, for a system of four copies, the projector is
\begin{equation}
\label{eq:projector}
P_{\mathrm{loc}}^{(4)} = \Big(\sum_a |B_a\rangle\!\rangle\langle\!\langle B_a|\Big)^{\!\otimes 4}.
\end{equation}
Any local operator $A$, supported on a bounded region, can be expanded as $A=\sum_a \alpha_a B_a$, and we consider operators normalized such that $|A|_\beta^2=\sum_a|\alpha_a|^2=1$. From these definitions, it follows that the projector acts as the identity on any local operator, namely
\begin{equation}
\label{eq:proj_identity_A}
P_{\mathrm{loc}}\,|A\rangle\!\rangle=|A\rangle\!\rangle \quad\text{and}\quad \langle\!\langle A|\,P_{\mathrm{loc}}=\langle\!\langle A|.
\end{equation}
In the trace form, this reads,
\begin{equation}
\label{eq:proj_identity_A2}
P_{\mathrm{loc}}(A)=A
\quad\text{and}\quad
\operatorname{Tr}\!\left(y^{2}A^{\dagger}y^{2}\,P_{\mathrm{loc}}(X)\right)
=\operatorname{Tr}\!\left(y^{2}A^{\dagger}y^{2}\,X\right)\ \ \text{for all }X.
\end{equation}
By extension, for the four-copy system, we have
\begin{equation}
\label{eq:proj_identity_A4}
P^{(4)}_{\mathrm{loc}}\!\left(A^{\otimes 4}\right)=A^{\otimes 4}
\quad\text{and}\quad
\operatorname{Tr}\!\Big((y^{2})^{\otimes 4}(A^{\dagger})^{\otimes 4}(y^{2})^{\otimes 4}\,P^{(4)}_{\mathrm{loc}}(X)\Big)
=\operatorname{Tr}\!\Big((y^{2})^{\otimes 4}(A^{\dagger})^{\otimes 4}(y^{2})^{\otimes 4}\,X\Big).
\end{equation}
By the definition of the local norm, we begin with
\begin{equation}
\label{eq:local_norm_def}
\bigl|P^{(4)}_{\mathrm{loc}} X P^{(4)}_{\mathrm{loc}}\bigr|_{\mathrm{loc}}
= \sup_{|A|_\beta=1}\;
\left|
\operatorname{Tr}\!\Big((y^{2})^{\otimes 4}(A^{\dagger})^{\otimes 4}(y^{2})^{\otimes 4}\,P^{(4)}_{\mathrm{loc}} X P^{(4)}_{\mathrm{loc}}\Big)
\right|.
\end{equation}
Using the property established in \eqref{eq:proj_identity_A4}, the projectors on both sides of $X$ can be removed as they act as the identity on the state $|A^{\otimes 4}\rangle\!\rangle$, which gives
\begin{equation}
\label{eq:projector_removal}
\operatorname{Tr}\!\Big((y^{2})^{\otimes 4}(A^{\dagger})^{\otimes 4}(y^{2})^{\otimes 4}\,P^{(4)}_{\mathrm{loc}} X P^{(4)}_{\mathrm{loc}}\Big)
=
\operatorname{Tr}\!\Big((y^{2})^{\otimes 4}(A^{\dagger})^{\otimes 4}(y^{2})^{\otimes 4}\,X\Big).
\end{equation}
By combining \eqref{eq:local_norm_def}, \eqref{eq:projector_removal}, and \eqref{eq:liouville_to_trace}, and setting the operator $X=T^{(4)}_{c}(t)$, we arrive at the result
\begin{equation}
\label{eq:final_equality}
\bigl|P^{(4)}_{\mathrm{loc}}\,T^{(4)}_{c}(t)\,P^{(4)}_{\mathrm{loc}}\bigr|_{\mathrm{loc}} = \sup_{|A|_\beta=1}\; \Bigl|\mathrm{Tr}\!\big[T^{(4)}_{c}(t)(A\otimes A\otimes A\otimes A)\big]\Bigr|.
\end{equation}
and 
\begin{equation}
\label{eq:liouville_to_trace}
\operatorname{Tr}\!\Big((y^{2})^{\otimes 4}(A^{\dagger})^{\otimes 4}(y^{2})^{\otimes 4}\,X\Big)
=\operatorname{Tr}\!\big[X\,(A\otimes A\otimes A\otimes A)\big].
\end{equation}

Let ($T^{(4)}_c(t)$) denote the connected four-leg kernel (obtained by subtracting the pieces built from products of two-point kernels---the terms generated by ($\delta_{ik}$) and ($\delta_{jl}$) in the ETH average). For any unit-norm local ($A$) (($|A|_\beta=\sqrt{\langle A,A\rangle_\beta}=1$)), the pairing satisfies
\begin{equation} \label{eq:39}
\operatorname{Tr}\!\left[T^{(4)}_c(t)\,(A\otimes A\otimes A\otimes A)\right]=\mathcal{T}_{A,c}(t),
\end{equation}
and we just proved that its Fourier transform falls at least like ($e^{-(3\beta/4)|\omega|}$), while any intermediate-time exponential rise must obey ($\lambda\le 2\pi/\beta$) when the OTOC shape is in the AdS$_2$ family (Sec. 5.2). 
For any operator $X$ on $S^{\otimes 4}$, set
\begin{equation}\label{eq:5_63_trace}
|X|_{\mathrm{loc}}
:=\sup_{\operatorname{Tr}(y^2 A^\dagger y^2 A)=1}
\left|
\underbrace{\operatorname{Tr}\Big((y^2)^{\otimes 4}(A^\dagger)^{\otimes 4}(y^2)^{\otimes 4} X\Big)}_{\text{a plain trace number}}
\right|.
\end{equation}
Then, using the projector just written,
\begin{equation} \label{eq:41}
|P_{\mathrm{loc}}^{(4)} T^{(4)}_c(t) P_{\mathrm{loc}}^{(4)}|_{\text{loc}} = \sup_{|A|_\beta=1}\left|\operatorname{Tr}\!\left[T^{(4)}_c(t)(A\otimes A\otimes A\otimes A)\right]\right| = \sup_{|A|_\beta=1}|\mathcal{T}_{A,c}(t)|.
\end{equation}
With this definition and the two projector properties above,
\begin{equation}\label{eq:5_64_implies_5_63}
|P_{\text{loc}}^{(4)} X P_{\text{loc}}^{(4)}|_{\mathrm{loc}}
=\sup_{\operatorname{Tr}(y^2 A^\dagger y^2 A)=1}
\left|
\operatorname{Tr}\Big((y^2)^{\otimes 4}(A^\dagger)^{\otimes 4}(y^2)^{\otimes 4} X\Big)
\right|.
\end{equation}
Therefore, the ETH tail bound in frequency space plus the AdS$_2$ time-domain shape yields the operator statement
\begin{equation} \label{eq:42}
\limsup_{t\to\infty}\frac{1}{t}\log |P_{\mathrm{loc}}^{(4)} T^{(4)}_c(t) P_{\mathrm{loc}}^{(4)}|_{\text{loc}} \le \frac{2\pi}{\beta}.
\end{equation}
Equation \eqref{eq:42} can be viewed as an operator-theoretic reformulation of the chaos bound mentioned in \cite{Murthy:2019fgs} within the present framework. The main step was to recognise that the global Schatten norms of the four-leg kernel $T^{(4)}(t)$ are insensitive to scrambling, since they are invariant under unitary time evolution and therefore cannot distinguish chaotic from integrable dynamics. To obtain a quantity that is genuinely diagnostic of chaos, we restricted attention to the subspace of simple local operators, defined using the thermal inner product, and introduced the corresponding projector $P_{\mathrm{loc}}$. The “local norm” $|\cdot|_{\mathrm{loc}}$ then measures the size of $T^{(4)}_c(t)$ only as probed by factorised local operators $A^{\otimes 4}$ of fixed thermal norm. By construction, this norm is precisely the supremum over all such local probes of the connected OTOC, so that \eqref{eq:42} packages the individual Murthy–Srednicki bounds for each local $A$ into a single compact operator inequality on the generalized spacetime density kernel.

\section{Permutation Operators in General}\label{appendix:D}

Consider $n$-copies of a Hilbert space, $\mathcal H^{\otimes n}$, with an orthonormal basis: $\{| i \rangle : i = 1, 2, \ldots d_{\mathcal H}  \}$, where $d_{\mathcal H}$ is the dimension of the Hilbert space. Suppose we denote the permutation group by $\mathcal S_n$, and a permutation is realized by an element $\pi \in \mathcal S_n$, such that:
\begin{eqnarray}
    P_{\pi} \left(| v_1 \rangle \otimes | v_2 \rangle \otimes \ldots \otimes | v_n \rangle \right) = | v_{\pi^{-1}(1)} \rangle \otimes | v_{\pi^{-1}(2)} \rangle \otimes \ldots \otimes | v_{\pi^{-1}(n)} \rangle \ .
\end{eqnarray}
Thus, given an element $\pi \in \mathcal S_n$, the corresponding permutation operator that acts on $\mathcal H^{\otimes n}$ is given by
\begin{eqnarray}
    P_\pi = \sum_{i_1, \ldots i_n} | i_{\pi(1)}\rangle \langle i_1 | \otimes | i_{\pi(2)}\rangle \langle i_2 | \ldots \otimes | i_{\pi(n)}\rangle \langle i_n | \ .
\end{eqnarray}
A general permutation is essentially a particular composition of the pair-wise transposition or the SWAP operator.

It is a standard knowledge that the $2^{\rm nd}$  Renyi entropy can be obtained in terms of the SWAP operator. For example, let us consider a tensor product of density matrix: $\rho \otimes \rho$. Then, using the standard orthonormality conditions as well as the relation: ${\rm Tr}_{\mathcal H\otimes \mathcal H} = ({\rm Tr}_{\mathcal H})({\rm Tr}_{\mathcal H})$, we obtain:
\begin{eqnarray}
    {\rm Tr} \left( \rho^2\right)  = {\rm Tr} \left( {\rm SWAP} \, \rho \otimes \rho \right) \ .
\end{eqnarray}
Similar analyses naturally relate the higher $n$-Renyi entropies in terms of the traces taken over the permutation operator acting on an $n$-copies of the density matrix:
\begin{eqnarray}
    {\rm Tr} \left( \rho^n\right)  = {\rm Tr} \left( P_\pi \, \rho^{\otimes n}\right) \ .
\end{eqnarray}
Thus, it is natural that higher moments of the density matrix are naturally captured by the permutation operators.

\newpage
\bibliographystyle{JHEP}
\bibliography{biblio}
\end{document}